\documentclass[aps,superscriptaddress,showpacs,twocolumn,preprintnumbers]{revtex4}
\usepackage{amsmath,amssymb,bm} \usepackage{graphicx}
\usepackage{psfrag,color}
\usepackage{hyperref}

\newcommand{\dRpa}{\langle\Delta R_{\parallel}\rangle}
\newcommand{\drho}{\langle\Delta\rho\rangle}
\newcommand{\Rpa}{R_{\parallel}}
\newcommand{\lpa}{\ell_{\parallel}}
\newcommand{\lp}{\ell_{\perp}}
\newcommand{\zet}{\hat{\zeta}}
\newcommand{\tlp}{t^{\parallel}_{L}}

\renewcommand{\vec}{\mathbf}

\begin{document}

\title{Longitudinal Response of Confined Semiflexible Polymers}

\author{Florian Th\"{u}roff}
	\affiliation{Arnold Sommerfeld Center for Theoretical Physics (ASC) and Center for NanoScience (CeNS), LMU M\"{u}nchen, Theresienstra{\ss}e 37, 80333 M\"{u}nchen, Germany}
\author{Benedikt Obermayer}
	\affiliation{Arnold Sommerfeld Center for Theoretical Physics (ASC) and Center for NanoScience (CeNS), LMU M\"{u}nchen, Theresienstra{\ss}e 37, 80333 M\"{u}nchen, Germany}
	\affiliation{Department of Physics, Harvard University, Cambridge, MA 02138, USA}
\author{Erwin Frey}
	\affiliation{Arnold Sommerfeld Center for Theoretical Physics (ASC) and Center for NanoScience (CeNS), LMU M\"{u}nchen, Theresienstra{\ss}e 37, 80333 M\"{u}nchen, Germany}

\begin{abstract}

The longitudinal response of single semiflexible polymers to sudden changes in externally applied forces is known to be controlled by the propagation and relaxation of backbone tension. Under many experimental circumstances, realized, \textit{e.g.}, in nano-fluidic devices or in polymeric networks or solutions, these polymers are effectively confined in a channel- or tube-like geometry. By means of heuristic scaling laws and rigorous analytical theory, we analyze the tension dynamics of confined semiflexible polymers for various generic experimental setups. It turns out that in contrast to the well-known linear response, the influence of confinement on the non-linear dynamics can largely be described as that of an effective prestress. We also study the free relaxation of an initially confined chain, finding a surprising superlinear $\sim t^{9/8}$ growth law for the change in end-to-end distance at short times.

\end{abstract}


\date{\today}

\preprint{LMU-ASC 68/10}

\maketitle

\section{\label{sec:i}Introduction}

Semiflexible polymers such as actin or microtubules play essential roles for the elastic behavior of cellular structures \cite{Bausch:2006p992,Fletcher:2010p1497}. Their mechanical properties are determined by their relatively high bending stiffness, characterized through a compared to the contour length $L$ large persistence length $\ell_p$. 
Many of the static and dynamic properties of single filaments have been investigated in the last decades, and many of the results carry over to more complex structures such as polymer networks \cite{Wilhelm:2003p649,Head:2003p933,Heussinger:2006p991,Bausch:2006p992} and solutions \cite{Harnau:1995p1494,Harnau:1996p1491,Isambert:1996p928,Morse:1998p924,Hinner:1998p964,Morse:2001p926,Dogic:2004p1482,Hinsch:2007p934,Semmrich:2007p998}, the latter of which are theoretically described through single filaments confined to an effective tube.
In recent years, much experimental research is also focused on single molecule studies, by means of powerful new nano-technological methods, where filaments are routinely confined to micro- or nanometer-sized channels \cite{Reisner:2005p1004,Tegenfeldt:2004p1146,Jo:2007p619,Bakajin:1998p1191,Koster:2005p943,Choi:2005p1190,Balducci:2007p1193,Koster:2008p940,Koester:2009p1508,Persson:2010p1511,Levy:2010p1543}.

Theoretical understanding of confined semiflexible polymers has mainly been gained in terms of equilibrium statistics \cite{Harnau:1999p1461,Bicout:2001p1149,Yang:2007p1148,Levi:2007p459,Wagner:2007p458,Cifra:2009p732,Cifra:2009p737,Thuroff:2010p1769,Burkhardt:2010p1479} and some dynamical properties in microcapillary flows \cite{Chelakkot:2010p14001} and on the linear response level \cite{Granek:1997p1170,Morse:1998p1276,Nam:2010p1691}. However, for larger forces such as those commonly applied in single-molecule experiments, it is important to account for the extremely high stretching stiffness of most semiflexible polymers, which as an effective inextensibility has not only significant consequences for the static force-extension relation but also heavily influences dynamical properties. The responsible nonlinearities are especially apparent when externally applied forces are suddenly changed, for in this case the backbone tension, which prevents stretching of monomer bonds, shows a highly nontrivial spatial and temporal dependence. Previous approaches to capture this phenomenon were based on heuristic scaling arguments and effective theories \cite{Seifert:1996p779,Ajdari:1997p1171,Everaers:1999p750,BrochardWyart:1999p798}. These early results were later confirmed and generalized by means of a systematic formalism based on a multiple scale perturbation theory \cite{Hallatschek:2005p226,Hallatschek:2007p222,Hallatschek:2007p229}. Recent extensions of these results  include prestressed filaments \cite{Obermayer:2007p643}, transverse forces \cite{Obermayer:2007p644}, extensible backbones \cite{Obermayer:2009p741}, and oscillatory forces \cite{Hiraiwa:2009p961}. A common observation of these studies is that the essential nonlinearities contained in the nontrivial tension profile lead to a mixing of short- and long-wavelength fluctuations, such that experimentally changing the equilibrium mode spectrum of contour fluctuations can have far more drastic consequences as usually suspected, for instance on the scenario-specific relaxation dynamics observed for polymers that were prepared in initially straight conformations by different means \cite{Obermayer:2009p740}.

\begin{figure}[b]
	\centering
		\includegraphics[width=.35\textwidth]{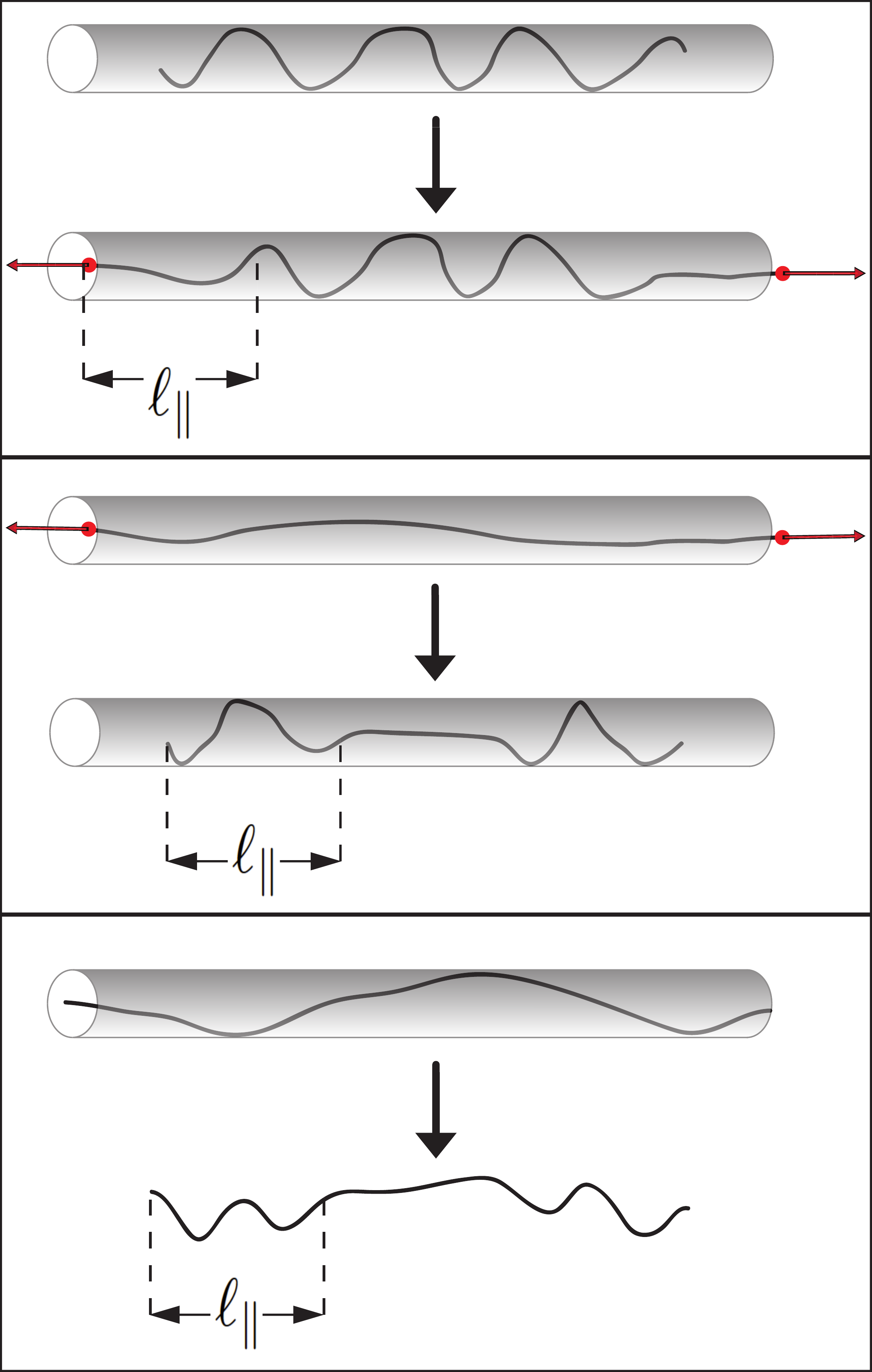}
	\caption{Illustration of the three scenarios investigated in this work. \textit{Top (pulling)}: Both ends of an initially tension free backbone, which is equilibrated in a channel of constant diameter, are pulled apart by an externally applied force $\mathfrak{f}$. \textit{Middle (release)}: The confined polymer is equilibrated under an externally applied prestretching force $\mathfrak{f}$, which is suddenly released. \textit{Bottom (free relaxation from confinement)}: The contour is equilibrated in confinement and relaxes in free space. In all cases, the sudden change in external conditions propagates along the contour within growing boundary layers of size $\lpa(t)$.}
	\label{fig:scenarios}
\end{figure}

It is well known that confinement leads to a changed fluctuation spectrum, because contour undulations with a wavelength longer than Odijk's deflection length $L_d\sim\ell_p^{1/3}D^{2/3}$ ($D$ denoting the channel diameter) \cite{Odijk:1983p211} are strongly suppressed. Under many circumstances,  this merely leads to a renormalization of some characteristic length and time scales \cite{Granek:1997p1170,Morse:1998p1276,KroyFrey}. However, in the light of the above we expect substantial changes for the dynamics of the tension along the contour. Because of the widespread presence of confinement in common single-molecule experiments, we analyze the effects of confinement on the longitudinal response of semiflexible polymers in three generic scenarios (cf. Fig. \ref{fig:scenarios}). First, we consider the sudden stretching of a semiflexible chain in cylindrical confinement upon applying a constant ``pulling'' force at its ends. Next, we discuss the inverse ``release'' scenario, where this stretching force is suddenly removed. Compared to the corresponding unconfined scenarios analyzed previously, we find in both cases dynamic signatures strongly reminiscent of a prestress. In order to clearly elucidate the specific dynamical consequences of the different contour statistics of a chain confined to a channel, we finally address the idealized scenario where this constraint is suddenly removed. We will refer to this latter scenario as ``free relaxation from confinement'' (FRC).

The paper is organized as follows. In the next section \ref{sec:e} we present the basic mathematical formulation of the problem and derive the equations of motion. The linearized version of these equations of motion are investigated on a scaling level in section \ref{sec:s}. There the physical principles behind all subsequently treated relaxation processes are illuminated and scaling laws governing the longitudinal dynamics are derived. Section \ref{sec:a} contains the analytical core of this work. After a brief survey of the ideas that lead Hallatschek \textit{et al.} \cite{Hallatschek:2005p226,Hallatschek:2007p222} to a refined formulation of the relaxation dynamics of semiflexible polymers, we adjust this description to the particular scenarios outlined above and investigate the tension dynamics within the various asymptotic limits. The following section \ref{sec:l} builds upon these analytical results to discuss the longitudinal dynamics of the polymer's projected end-to-end distance. Section \ref{sec:experiments} quantifies our analytical findings in terms of numerical estimates for time, force, and length scales relevant in experiments. Finally section \ref{sec:su} summarizes the main results of this work.

\section{\label{sec:e}Equations of Motion}
To determine the polymer's equations of motion, we resort to the wormlike chain model \cite{KRATKY:1949p1000} in its continuous version \cite{SAITO:1967p1001}. According to this model the polymer is described by means of an inextensible continuous space curve $\vec{r}(s)$ and a bending Hamiltonian $\mathcal{H}_{\text{bnd}}=\kappa/2\,\int ds\,\vec{r}''^2$ introducing an energetic penalty proportional to the chain's (squared) curvature $\vec{r}''^2$ to account for bending rigidity (primes indicate derivatives with respect to the arc length coordinate $s$). In three dimensions the modulus $\kappa$ is related to the persistence length $\ell_p$, which characterizes the decay of tangent-tangent correlations along the contour, via $\kappa=k_BT\ell_p$. Confinement in a hard-walled channel will be modeled by means of an additional harmonic potential $\mathcal{H}_{\text{pot}}=\gamma/2\,\int ds\,\vec{r}_{\perp}^2$ punishing fluctuations $\vec{r}_{\perp}$ away from the symmetry axis of the channel. The relation between the effective confinement strength $\gamma$ and the diameter $D$ of the corresponding hard-walled channel is discussed in section \ref{sec:experiments} (cf. Ref. \cite{Yang:2007p1148}).
Further introducing the backbone tension $f(s,t)$ as a Lagrange multiplier function ensuring local inextensibility of the chain ($\vec{r}'(s)^2=1$) \cite{GOLDSTEIN:1995p1003}, the Hamiltonian reads
\begin{equation}
\label{eq:Hamiltonian}
\mathcal{H}=\frac{1}{2}\int_0^Lds\left[\kappa\vec{r}''(s)^2+f(s)\vec{r}'(s)^2+\gamma\vec{r}_{\perp}(s)^2\right].
\end{equation}
The polymer's overdamped motion thus follows from equating friction forces on the one hand and elastic and thermal forces on the other hand:
\begin{equation}
\label{eq:forcebalance}
\bm{\zeta}\partial_t\vec{r}(s,t)=\frac{\delta}{\delta\vec{r}(s,t)}\,\mathcal{H}\left[\vec{r}\right]+\bm{\xi}(s,t).
\end{equation}
The slender shape of the polymer allows to decompose the friction matrix $\bm{\zeta}=\zeta_{\perp}[(1-\vec{r}'\otimes\vec{r}')+\zet\,\vec{r}'\otimes\vec{r}']$ into a parallel and transverse component \cite{DoiEdwards}, where the parameter $\zet\approx1/2$ quantifies the friction anisotropy.

In order to simplify notation, we shall set $\kappa=1$ and $\zeta_{\perp}=1$ from now on, making forces a length$^{-2}$, times a length$^{4}$ and the effective confinement strength $\gamma$ a length$^{-4}$. Now, assuming the chain to be either strongly confined or sufficiently stiff, the transverse gradients $\vec{r}_{\perp}'^2=\mathcal{O}(\epsilon)\ll1$ become very small (we will give a definition of $\epsilon$ in terms of the system's parameters below). Moreover, choosing the $z$-axis to coincide with the direction of the polymer's extension allows us to effectively discriminate between transverse and longitudinal displacements and to write $\vec{r}(s)=(\vec{r}_{\perp}(s),s-r_{\parallel}(s))^t$, where $r_{\parallel}$ quantifies the amount of contour length stored in thermal undulations. For weakly bending polymers the inextensibility constraint $\vec{r}'^2=1$ implies $r_{\parallel}'=\vec{r}_{\perp}'^2/2+o(\epsilon)$ and we may expand Eq. \eqref{eq:forcebalance} in powers of $\epsilon$.
Keeping terms up to order $\mathcal{O}(\epsilon)$, we thus arrive at the following equations of motion, governing the chain's transverse and longitudinal dynamics:
\begin{subequations}
\label{eq:EM}
\begin{eqnarray}
\partial_t\vec{r}_{\perp}		&=& -\vec{r}_{\perp}''''-\gamma\vec{r}_{\perp}+[f\vec{r}_{\perp}']'+\bm{\xi}_{\perp}\\
\zet\partial_tr_{\parallel} &=& (\zet-1)\vec{r}_{\perp}'^{t}\partial_t\vec{r}_{\perp}-r_{\parallel}''''-f'+\left(fr_{\parallel}'\right)'+\xi_{\parallel}.
\end{eqnarray}
\end{subequations}
Solving Eqs. \eqref{eq:EM} by means of ordinary perturbation theory only works for large enough times, but runs into trouble in the limit of short times. We will thus present a refined perturbation theory in section \ref{sec:a}.

\section{\label{sec:s}Scaling Picture}
\subsection{Length scales}

Much of the physics underlying the relaxation mechanisms of confined semiflexible chains can be understood by considering the interplay between the various characteristic length scales. Apart from the contour length $L$ and the persistence length $\ell_p$, four additional length scales can be identified, which are crucial to the dynamics of such polymers (see table \ref{tab:lengthscales} for a summary).

Two static length scales may readily be inferred from the externally applied force $\mathfrak{f}$ and the confinement strength $\gamma$. A simple dimensional analysis reveals that, in our system of units, the length $\gamma^{-1/4}$ may be identified with Odijk's deflection length \cite{Odijk:1983p211}, whereas the scale $\mathfrak{f}^{-1/2}$, set by the externally applied force, is nothing else but the very length for which $\mathfrak{f}$ represents the critical Euler load. We shall refer to this latter scale as ``Euler buckling length''.

Apart from these static scales, two characteristic dynamical length scales have been identified to play a key role in the understanding of the polymer's relaxation dynamics, the first of which can easily be deduced using the linearized version of the equations of motion \eqref{eq:EM}
\begin{equation}
\label{eq:linEM}
\partial_t\vec{r}_{\perp}=-\vec{r}_{\perp}''''-\gamma\vec{r}_{\perp}+f\vec{r}_{\perp}''+\bm{\xi}_{\perp}+o(\epsilon^{1/2}).
\end{equation}
Here we used
\begin{equation}
\label{eq:consttension}
f'=0
\end{equation}
(to lowest order) from the expansion of the longitudinal part of \eqref{eq:EM}. Denoting the characteristic length scale for transverse fluctuations by $\lp$, inspection of Eq. \eqref{eq:linEM} on a scaling level
\begin{equation}
\label{eq:linEMscaling}
t^{-1}\sim\lp^{-4}+\mathfrak{f}\,\lp^{-2}+\gamma
\end{equation}
(where the externally applied force $\mathfrak{f}$ sets the scale for the backbone tension $f$) immediately yields the scaling law for $\lp$
\begin{equation}
\label{eq:lpscaling}
\lp(t)\sim
\begin{cases}
t^{1/4}, & t\ll\min\{\mathfrak{f}^{-2},\gamma^{-1}\}\\
\sqrt{\mathfrak{f}\,t}, & \mathfrak{f}^{-2}\ll t\ll\gamma^{-1}.
\end{cases}
\end{equation}
As will be obvious from a more detailed treatment of the linearized equation of motion \eqref{eq:linEM} in section \ref{sec:a}, at time $t$ the transverse fluctuations of the polymer are in equilibrium with their surroundings on length scales $\sim\lp(t)$, which is why we refer to $\lp$ as (transverse) ``equilibration length''. Note, in particular, that the second line of Eq. \eqref{eq:lpscaling} only applies in cases where the Euler buckling length is much smaller than Odijk's deflection length. 

This last statement can be re-expressed in more intuitive terms. A simple dimensional analysis reveals that confinement induces a new force scale $\gamma^{1/2}$ which we will refer to as ``effective confinement force''. The concept of identifying confinement as an effective force will be central in the further course of our discussion. In particular, it allows to directly compare the strengths of applied forces and confinement. To this end, we introduce the ``force scale separation parameter''
\begin{equation}
\label{eq:x}
x\equiv\frac{\gamma^{1/2}}{\mathfrak{f}},
\end{equation}
measuring the effective confinement force $\gamma^{1/2}$ in units of the externally applied force $\mathfrak{f}$. So for the second scaling regime in Eq. \eqref{eq:lpscaling} to occur, the effective confinement force $\gamma^{1/2}$ has to be small compared to the externally applied force $\mathfrak{f}$, viz. $x\ll1$.

\begin{table}
	\begin{ruledtabular}
		\begin{tabular}{c|l}
			$\gamma^{-1/4}$       & Odijk's deflection length\\
			$\mathfrak{f}^{-1/2}$ & Euler buckling length\\
			$\lp(t)$              & (transverse) equilibration length\\
			$\lpa(t)$             & boundary layer size\\
		\end{tabular}
	\end{ruledtabular}
	\caption{Summary of characteristic length scales}
	\label{tab:lengthscales}
\end{table}

In the remainder of this contribution we shall mainly focus on the case $x\ll1$, where the chain's response to a large external perturbation is non-linear, discussing the case $x\gg1$ of comparably small external perturbations along the way. The experimental accessibility of these limits will be investigated in section \ref{sec:experiments}.

According to the ordinary perturbation scheme employed to arrive at the linearized equation of motion \eqref{eq:linEM} $\lp$ would be the only problem-specific dynamical length scale. As has been recognized before \cite{Seifert:1996p779,Ajdari:1997p1171,Everaers:1999p750,Pasquali:2001p1172}, however, ordinary perturbation theory (OPT) breaks down in the limit $t\rightarrow0$, where longitudinal friction becomes important. As can be seen from Eq. \eqref{eq:consttension}, the effect of longitudinal friction is of higher order in $\epsilon$ and thus absent in leading order OPT. In order to account for the effect of longitudinal friction, at least on a heuristic level, the statement of a spatially constant tension, Eq. \eqref{eq:consttension}, has to be reconsidered. In the short time limit, longitudinal friction prevents the bulk of the polymer from responding to sudden changes in ambient conditions and thus confines the longitudinal dynamics to boundary layers of dynamical size $\lpa(t)$ (cf. Fig. \ref{fig:scenarios}). Hence applicability of Eq. \eqref{eq:consttension} has to be restricted to local scales of size $\lp(t)$, with tension variations taking place at the much larger length scale $\lpa(t)$. This separation of scales is the basis for a more sophisticated multiple scale perturbation theory proposed in Refs. \cite{Hallatschek:2005p226,Hallatschek:2007p222}, which will be briefly reviewed in section \ref{sec:a}.
 
Before we go on to discuss the scaling behavior of the chain's longitudinal response, we briefly introduce one further central quantity in the discussion of the longitudinal dynamics of bio-polymers: the density of contour length stored in thermal undulations  $\langle\rho\rangle=\langle r_{\parallel}'\rangle\approx\frac{1}{2}\langle\vec{r}_{\perp}'^2\rangle$, where the last approximation holds for weakly bending chains. Since we may apply equilibrium theory on scales $\sim\lp(t)$, for the time being we are only interested in the equilibrium value $\langle\rho\rangle_0$ of this quantity, which we refer to as ``stored length density'' in what follows. To find an explicit expression for the spatial average of $\langle\rho\rangle_0$ in the weakly bending limit, we expand the transverse displacements in simple sine-modes \cite{Granek:1997p1170} $\vec{r}_{\perp}(s)=\sqrt{2/L}\sum_n\vec{a}_n\sin(n\pi s/L)$. Noting that the stored length density reads $\langle\rho\rangle_0=(2L)^{-1}\sum_n(n\pi/L)^2\langle\vec{a}_n^2\rangle$, where the  fluctuation amplitude $\langle\vec{a}_n^2\rangle$ can be determined by equipartition of energy, we find
\begin{equation}
\label{eq:initmodespectrum}
\langle\rho\rangle_0=\frac{1}{L\ell_p}\sum_{n=1}^{\infty}\frac{(\frac{n\pi}{L})^2}{(\frac{n\pi}{L})^4+\mathfrak{f}(\frac{n\pi}{L})^2+\gamma}.
\end{equation}
Depending on the magnitude of tension and confinement, this corresponds to the well known limits for the stored length density \cite{KOVAC:1982p1194,Odijk:1983p211,Marko:1995p1041,MACKINTOSH:1995p931}:
\begin{equation}
\label{eq:epsilondef}
\langle\rho\rangle_0\simeq
\begin{cases}
\frac{L}{6\ell_p} & \max\{\mathfrak{f},\gamma^{1/2}\}\ll L^{-2}\\
\frac{1}{2\ell_p\sqrt{\mathfrak{f}}} & \max\{\gamma^{1/2},L^{-2}\}\ll\mathfrak{f}\\
\frac{1}{2\ell_p\sqrt{2\gamma^{1/2}}} & \max\{\mathfrak{f},L^{-2}\}\ll\gamma^{1/2}.
\end{cases}
\end{equation}
Comparison of the stored length density for force- and confinement-dominated situations (\textit{i.e.} the second and third line of Eq. \eqref{eq:epsilondef}) suggests to identify effective confinement force and externally applied force (or more exactly $2\gamma^{1/2}\leftrightarrow\mathfrak{f}$). We will see in subsequent sections that this conclusion carries over to non-equilibrium situations, where the dynamical scaling laws characterizing the relaxation processes for both force scenarios (pulling and release) can be identified with those of prestretched chains (cf. Ref. \cite{Obermayer:2007p643}). Most prominently we will re-encounter the effective confinement force in the scenario of free relaxation from confinement, where an initially tension free contour builds up a bulk-tension which is equal to $2\gamma^{1/2}$ during relaxation in free space. Moreover, Eq. \eqref{eq:epsilondef} lends itself to define the small parameter $\epsilon$ quantifying the strength of transverse fluctuations, which was heuristically introduced in the last section. Recalling that $\langle\rho\rangle_0\sim\langle\vec{r}_{\perp}'^2\rangle\sim\epsilon$, we find $\epsilon\in\{L\ell_p^{-1},\mathfrak{f}^{-1/2}\ell_p^{-1},\gamma^{-1/4}\ell_p^{-1}\}$, depending on whether one is dealing with free, strongly prestretched or strongly confined chains, respectively. Note that the two latter definitions of $\epsilon$ allow for the use of the weakly bending rod approximation even for quite flexible polymers, such as dsDNA, provided the prestress or confinement is sufficiently strong.

\subsection{Longitudinal response}

Having identified all characteristic length scales, we go on to illustrate the physical mechanisms governing the relaxation dynamics of harmonically confined semiflexible chains within the short time tension propagation regime $t\ll\tlp$. Here the time scale $\tlp$ marks the time when the boundary layers eventually span over the whole contour, which is implicitly defined via $\lpa(\tlp)=L$. To this end we derive the scaling laws for the change in projected end-to-end distance
\begin{equation}
\label{eq:dRpa}
\dRpa(t)\equiv\langle\Rpa(t)-\Rpa(0)\rangle.
\end{equation}
These results will be confirmed in the course of a more rigorous analytical treatment of the underlying equations of motion in section \ref{sec:a}. The arguments leading to the scaling picture are analogous in all three cases, whence we will confine ourselves to the discussion of the pulling scenario.

In order to arrive at a scaling description for the chain's longitudinal response $\dRpa$ we follow the reasoning in Ref. \cite{Obermayer:2007p643}. The boundary layers of size $\lpa$, to which the longitudinal dynamics is confined, may notionally be subdivided into much smaller segments of length $\lp\ll\lpa$, where the polymer's transverse fluctuations are in equilibrium with their respective surroundings. Each boundary layer is hence composed of $\lpa/\lp$ segments and the scaling for $\dRpa$ may be inferred via the accumulated longitudinal response
\begin{equation}
\label{eq:dRpascaling}
\dRpa(t)\sim\frac{\lpa(t)}{\lp(t)}\,\delta,
\end{equation}
where the extension of each subsection $\delta$ follows from equilibrium theory. The scaling for $\lpa$ is determined by the observation that the force necessary to drag a polymer section of length $\lpa$ through the solvent must not exceed the externally applied force $\mathfrak{f}$ \cite{Everaers:1999p750}. In mathematical terms this condition translates to
\begin{equation}
\label{eq:lpascaling}
\zet\,\lpa(t)\,\frac{\lpa(t)}{\lp(t)}\,\frac{\delta}{t}\sim\mathfrak{f},
\end{equation}
which may be inverted to yield the scaling of $\lpa$.

Considering the pulling scenario, several asymptotic time regimes may be distinguished, depending on the magnitude of the equilibration length $\lp$ relative to the characteristic static length scales in the problem. For $x\ll1$ the Euler buckling length $\mathfrak{f}^{-1/2}$ is much smaller than Odijk's deflection length $\gamma^{-1/4}$, whence three distinct asymptotic time regimes emerge within the tension propagation regime:

\begin{enumerate}
\item For times $t\ll\mathfrak{f}^{-2}$, the polymer is equilibrated on length scales much shorter than Euler buckling length ($\lp\ll\mathfrak{f}^{-1/2}$). On local scales, the externally applied force $\mathfrak{f}$ may thus be considered as small perturbation and the chain relaxes freely ($\lp\sim t^{1/4}$). The extension per segment  $\delta\sim(\lp^4/\ell_p)\,\mathfrak{f}$ thus follows from the linearized force-extension relation \cite{MACKINTOSH:1995p931}.
\item For intermediate times $\mathfrak{f}^{-2}\ll t\ll\gamma^{-1/2}\mathfrak{f}^{-1}$, the equilibration length clearly exceeds the Euler buckling length, but is still small compared to Odijk's deflection length ($\mathfrak{f}^{-1/2}\ll\lp\ll\gamma^{-1/4}$). Within this time regime the chain is yet not influenced by confinement on local scales. The externally applied force, however, may be regarded as large and the relaxation behavior changes from free to forced relaxation ($\lp\sim\sqrt{\mathfrak{f}\,t}$). The small segments almost get completely stretched and we may estimate the extension per segment $\delta\sim\lp^2/\ell_p$ by assuming all initially stored length to be stretched out completely.
\item Finally, once the equilibration length exceeds Odijk's deflection length at times $\gamma^{-1/2}\mathfrak{f}^{-1}\ll t\ll\tlp$, confinement suppresses long wavelength modes in the initial spectrum of the stored length density. Again, assuming this stored length to be entirely pulled out by virtue of the large force $\mathfrak{f}$ we arrive at $\delta\sim(\gamma^{-1/4}/\ell_p)\,\lp$.
\end{enumerate}

Having determined the local extension $\delta$ within the various asymptotic time regimes, the scaling law for $\lpa$ may readily be obtained by means of equation \eqref{eq:lpascaling}.
\begin{equation}
\label{eq:lpascalingpull}
\lpa\sim\sqrt{\frac{\ell_p}{\zet}}\cdot
\begin{cases}
t^{1/8} & t\ll\mathfrak{f}^{-2}\\
(\mathfrak{f}\,t)^{1/4} & \mathfrak{f}^{-2}\ll t\ll\frac{1}{\sqrt{\gamma\mathfrak{f}^{2}}}\\
\gamma^{1/8}\sqrt{\mathfrak{f}\,t} & \frac{1}{\sqrt{\gamma\mathfrak{f}^{2}}}\ll t\ll\tlp
\end{cases}
\end{equation}

Hence, using Eq. \eqref{eq:dRpascaling} we find
\begin{equation}
\label{dRpascalingpull}
\dRpa\sim\frac{1}{\sqrt{\zet\ell_p}}\cdot
\begin{cases}
\mathfrak{f}\,t^{7/8} & t\ll\mathfrak{f}^{-2}\\
(\mathfrak{f}\,t)^{3/4} & \mathfrak{f}^{-2}\ll t\ll\frac{1}{\sqrt{\gamma\mathfrak{f}^{2}}}\\
\gamma^{-1/8}\sqrt{\mathfrak{f}\,t} & \frac{1}{\sqrt{\gamma\mathfrak{f}^{2}}}\ll t\ll\tlp
\end{cases}
\end{equation}

It is intuitively clear that these scaling relations are identical to the pulling scenario for unconfined chains as long as $\lp\ll\gamma^{-1/4}$, \textit{i.e.} as long as the chain has no perception of the confining channel walls on equilibrated length scales. In fact, the two time regimes for $t\ll\gamma^{-1/2}\mathfrak{f}^{-1}$ have already been discussed in the context of unconfined pulling in Refs. \cite{Everaers:1999p750,Morse:1998p924,Seifert:1996p779,Hallatschek:2005p226,Hallatschek:2007p229}, where identical scaling relations for $\lpa$ and $\dRpa$ have been discovered. In addition, analogous to the case of pulling chains subject to a prestretching force $f_{\text{pre}}$ \cite{Obermayer:2007p643}, confinement introduces an extra asymptotic time regime ($\gamma^{-1/2}\mathfrak{f}^{-1}\ll t\ll\tlp$) whose respective scaling relations are identical to those of prestretched pulling upon identifying $f_{\text{pre}}\leftrightarrow\gamma^{1/2}$. In this context confinement may thus be interpreted as effective prestretching force of magnitude $\sim\gamma^{1/2}$.

Before we go on to a more rigorous treatment of the equations of motion \eqref{eq:PIDE}, we briefly address the case $x\gg1$, \textit{i.e.} $\mathfrak{f}\ll\gamma^{1/2}$. Noting that, on a scaling level, the confined polymer may be regarded as being composed of $L/\gamma^{-1/4}$ independent chain segments, the above statement $x\gg1$ means that these segments are very stiff compared to the small externally applied force $\mathfrak{f}$, whence the whole chain responds linearly to the external perturbation. Pulling and release may thus be regarded as mutually inverse scenarios. The Euler buckling length $\mathfrak{f}^{-1/2}$ now exceeds Odijk's deflection length $\gamma^{-1/4}$ and only two time regimes may be distinguished during tension propagation, depending on whether $\lp\ll\gamma^{-1/4}$ for times $t\ll\gamma^{-1}$ or $\lp\gg\gamma^{-1/4}$ for times $t\gg\gamma^{-1}$. The situations for short times $t\ll\gamma^{-1}$ as well as the respective scaling laws are identical to those for $x\ll1$. The scaling laws for $t\gg\gamma^{-1}$ may be obtained by noting that the local extension $\delta\sim(\lp/\ell_p)\,\gamma^{-3/4}\mathfrak{f}$ follows from the linearized force-extension relation in confinement \cite{Wang:2007p1144}. Since the chain relaxes ``tension-free'' all the time ($\lp\sim t^{1/4}$), we arrive at the following scaling laws for $\dRpa$
\begin{equation}
\label{eq:dRpascalingpullxgg1}
\dRpa\sim\frac{\pm\mathfrak{f}}{\sqrt{\zet\ell_p}}\cdot
\begin{cases}
t^{7/8} & t\ll\gamma^{-1}\\
\gamma^{-3/8}\,t^{1/2} & t\gg\gamma^{-1}.
\end{cases}
\end{equation}
Here the plus sign refers to pulling, the minus sign to release. In Eq. \eqref{eq:dRpascalingpullxgg1} we recover the well-known $\sim t^{7/8}$ scaling law for short times and obtain a $\sim t^{1/2}$ scaling law for long times. Both scaling exponents have recently been deduced by Nam \textit{et al.} \cite{Nam:2010p1691} for confined chains within linear response theory.

\section{\label{sec:a}Asymptotic Equations of Motion}

\begin{table}
	\begin{ruledtabular}
		\begin{tabular}{c|c c c}
			  & Pulling & Release & FRC\\
			\hline
			$a$          & $0$     & $1$     & $0$                 \\
			$b$          & $x^2$   & $x^2$   & $1$                 \\
			$c_0$        & $\tau$  & $0$     & $0$                 \\
			$c_{\infty}$ & $0$     & $\tau$  & $0$                 \\
			$D$					 & $1$     & $1$     & $\partial_{\sigma}$
		\end{tabular}
	\end{ruledtabular}
	\caption{\label{tab:parameters}Summary of the scenario specific parameters occurring in Eqs. \eqref{eq:PIDErescaled} and \eqref{eq:BC}.}
\end{table}

\subsection{Linearized dynamics}
Even in the limit of short times, where the action of longitudinal friction causes the breakdown of OPT, the linearized equations of motion \eqref{eq:linEM} remain applicable at local scales of size $\sim\lp(t)$. A formal solution of Eq. \eqref{eq:linEM} is therefore the starting point for all subsequent discussions.

Eq. \eqref{eq:linEM} may readily be solved by means of a Green's function method. Concentrating on bulk dynamics we set $L\rightarrow\infty$ and determine the Green's function in Fourier space:
\begin{equation}
\label{eq:green}
\chi_{\perp}(q;t,\tilde{t})=\Theta(t-\tilde{t})\,e^{-q^2\left[q^2(t-\tilde{t})+F(t)-F(\tilde{t})\right]-\gamma(t-\tilde{t})},
\end{equation}
where 
\begin{equation}
\label{eq:F}
F(t)\equiv\int_0^td\hat{t}f(\hat{t})
\end{equation}
denotes the time integrated tension. Strictly speaking $\chi_{\perp}$ stated in Eq. \eqref{eq:green} would have to be complemented correspondingly, in order to account for problem specific boundary conditions. As discussed in Ref. \cite{Hallatschek:2007p222}, however, the impact of boundary conditions is of minor importance for our present purposes, which is why we will dispense with any further discussion of this issue, referring the interested reader to Ref. \cite{Obermayer:2009p740}. Given the response function $\chi_{\perp}$ in Eq. \eqref{eq:green}, mediating the influence of external and thermal forces, the mode spectrum of transverse displacements immediately follows by
\begin{equation}
\label{eq:rp}
\vec{r}_{\perp}(q,t)=\int_{-\infty}^{t}d\tilde{t}\chi_{\perp}(q;t,\tilde{t})\bm{\xi}(q,\tilde{t}). 
\end{equation}
Later on we will need the coarse grained stored length density\footnote{Coarse graining is realized by a spatial average over intermediate length scales which can be replaced by an ensemble average safe for subdominant contributions that depend on the boundary conditions for the contour \cite{Hallatschek:2007p222,Hallatschek:2007p229,Obermayer:2009p740}.} $\langle\rho\rangle=\langle\vec{r}_{\perp}'^2\rangle/2+o(\epsilon)$, which immediately follows via Eq. \eqref{eq:rp}:
\begin{equation}
\label{eq:rho}
\begin{split}
\langle\rho\rangle(t)=\int_0^{\infty}\frac{dq}{\pi\ell_p}&\Biggl\{\langle\rho\rangle_0(q)\chi_{\perp}(q;t,0)\\
                                                         &+2q^2\int_0^td\tilde{t}\,\chi_{\perp}(q;t,\tilde{t})\Biggr\},
\end{split}
\end{equation}
where we used $\langle\bm{\xi}_{\perp}(s,t)\bm{\xi}_{\perp}^t(\tilde{s},\tilde{t})\rangle=4/\ell_p\,\delta(s-\tilde{s})\delta(t-\tilde{t})$. Here the initial mode spectrum $\langle\rho\rangle_0$ may be inferred from a continuum limit of Eq. \eqref{eq:initmodespectrum}.

\subsection{Multiple scale perturbation theory (MSPT)}

As briefly discussed in section \ref{sec:s}, the notion of a constant tension, as expressed by equation \eqref{eq:consttension}, has to be reconsidered on scales beyond the equilibration length $\lp(t)$. On the much larger scales $\sim\lpa(t)$ longitudinal friction causes the build-up of a non-trivial tension profile which in turn dominates the global relaxation dynamics of the chain. In order to understand these connections quantitatively, OPT from above has to be refined to take into account large scale tension variations.
The key observation, leading to an improved description of relaxation dynamics suggested in Refs. \cite{Hallatschek:2005p226,Hallatschek:2007p222}, lies in the large separation between the typical scales characterizing transverse ($\lp$) and longitudinal ($\lpa$) dynamics, holding true for weakly bending chains. This locally allows for the use of the linearized equations of motion. Having at hand the respective scaling laws governing $\lp$ and $\lpa$ (cf. section \ref{sec:s}), it is a straightforward matter to check that $\lp/\lpa\lesssim\epsilon^{1/2}\ll1$ indeed holds in any of the cases considered in this work. As detailed in Refs. \cite{Hallatschek:2005p226,Hallatschek:2007p222}, this observation may be exploited to conduct a multiple scale perturbation analysis, which eventually relates the curvature of the local tension profile to the change in the coarse grained stored length density:
\begin{equation}
\label{eq:tensionstoredlength}
\partial^2_sF(s,t)=-\zet\langle\drho(s,t),
\end{equation}
where 
\begin{equation}
\label{eq:defdrho}
\drho(s,t)\equiv\langle\rho\rangle(s,t)-\langle\rho\rangle(s,0).
\end{equation}
In Eq. \eqref{eq:tensionstoredlength} the stored length density $\langle\rho\rangle$ inherits its arc length dependence adiabatically from the (time integrated) backbone tension $F$. Using Eqs. \eqref{eq:green}, \eqref{eq:rho} and \eqref{eq:tensionstoredlength} we thus arrive at the following closed relation all of our subsequent work is based on:
\begin{equation}
\label{eq:PIDE}
\begin{split}
&\partial^2_sF(s,t)=\\
& \zet\int_0^{\infty}\frac{dq}{\pi\ell_p}\Biggl\{\langle\rho\rangle_0(q)\left(1-e^{-2q^2\left[q^2t+F(s,t)\right]-2\gamma t}\right)\\
&- 2q^2\int_0^td\tilde{t}\,e^{-2q^2\left[q^2(t-\tilde{t})+F(s,t)-F(s,\tilde{t})\right]-2\gamma(t-\tilde{t})}\Biggr\}
\end{split}
\end{equation}

In order to corroborate the scaling picture drawn in the previous section we now switch on to a rigorous analysis of Eq. \eqref{eq:PIDE}. For the sake of convenience we introduce dimensionless variables which are defined by means of the following rescaling scheme:
\begin{subequations}
\begin{eqnarray}
\label{eq:rescale}
s &\rightarrow& \sigma\cdot\sqrt{\frac{\ell_p}{\zet}}\mathfrak{f}^{-1/4}\\
t &\rightarrow& \tau\cdot\mathfrak{f}^{-2}\\
q &\rightarrow& q\cdot\mathfrak{f}^{1/2}\\
F(s,t) &\rightarrow& \phi(\sigma,\tau)\cdot\mathfrak{f}^{-1}.
\end{eqnarray}
\end{subequations}
Here $\mathfrak{f}$, denoting the externally applied force in pulling and release setups, needs to be replaced by $\gamma^{1/2}$ in the case of free relaxation from confinement. Applying this rescaling procedure renders Eq. \eqref{eq:PIDE} in the form
\begin{equation}
\label{eq:PIDErescaled}
\begin{split}
\partial^2_{\sigma}\phi(\sigma,\tau) &= \int_0^{\infty}\frac{dq}{\pi}\Biggl\{\frac{q^2\left(1-e^{-2q^2\left[q^2\tau+\phi(\sigma,\tau)\right]-2b\tau}\right)}{q^4+aq^2+b}\\
                   &- 2q^2\int_0^{\tau}d\tilde{\tau}\,e^{-2q^2\left[q^2(\tau-\tilde{\tau})+\phi(\sigma,\tau)-\phi(\sigma,\tilde{\tau})\right]-2b(\tau-\tilde{\tau})}\Biggr\},
\end{split}
\end{equation}
where the scenario specific values of the parameters $a$ and $b$ are gathered in table \ref{tab:parameters}. All of our subsequent calculations are based on (intermediate) asymptotic expansions of Eq. \eqref{eq:PIDErescaled}, assuming the time integrated backbone tension $\phi$ to attain the following one parameter scaling form
\begin{equation}
\label{eq:scalingformphi}
\phi(\sigma,\tau)\simeq\tau^{\alpha+1}\hat{\phi}\left(\xi\equiv\frac{\sigma}{\tau^{\eta}}\right),
\end{equation}
with scaling exponents $\alpha$ and $\eta$ depending on the particular asymptotic time regime under consideration. The validity of this scaling assumption will be confirmed in the following subsections, where the values of $\alpha$ and $\eta$ are explicitly calculated. We also note in advance that the applicability of our subsequently developed approximation schemes relies on the ad hoc condition 
\begin{equation}
\label{eq:alpha}
\alpha>-1/2,
\end{equation}
which will indeed be justified with the benefit of hindsight. We complete our mathematical description of the system by stating the following scenario specific boundary conditions for semi-infinite filaments, supplementing Eq. \eqref{eq:PIDErescaled}:
\begin{subequations}
\label{eq:BC}
\begin{eqnarray}
\label{eq:BC1}
\phi(0,\tau) &=& c_0\\
\label{eq:BC2}
\lim_{\sigma\rightarrow\infty}\,D\,\phi(\sigma,\tau) &=& c_{\infty}.
\end{eqnarray}
\end{subequations}
Here the scenario specific constants $c_0$ and $c_{\infty}$ and the operator $D$ are given in table \ref{tab:parameters}. The assumption of dealing with semi-infinite contours amounts to concentrating on the dynamics at one of the polymer's ends only. This is possible since, during the tension propagation regime, the dynamics entirely occurs within boundary layers of size $\lpa(t)\ll L$ which are separated by a large bulk section and thus are independent from each other. For later times, of course (\textit{i.e.} when $\lpa(t)$ is comparable to the contour length $L$), this assumption has to be reconsidered.

\subsection{Pulling}

Adjusting Eq. \eqref{eq:PIDErescaled} to the pulling scenario (\textit{i.e.} setting $a=0$, $b=x^2\ll1$) we seek possible asymptotic expansions for the various time regimes identified in section \ref{sec:s}. Here the boundary conditions, stated in Eq. \eqref{eq:BC}, reflect the action of a constant pulling force $\mathfrak{f}$ at the ends, and a vanishing backbone tension in the bulk of the chain.

\subsubsection{\label{sec:pullshort}Short times}
We start our discussions with asymptotically short times $\tau\ll1$. Invoking our general scaling assumption Eq. \eqref{eq:scalingformphi} and assuming $\alpha>-1/2$, we conclude that the bending modes, which are of order $\mathcal{O}(q^4\tau)$ dominate over tension modes $\mathcal{O}(q^2\tau^{\alpha+1})$ in the exponentials. This allows us expand the exponentials up to first order in the tension modes. Since, moreover, we currently concentrate on cases with $x\ll1$, we may set $e^{-2x^2\tau}\approx1\approx e^{x^2(\tau-\tilde{\tau})}$, arriving at
\begin{equation}
\label{eq:pullshort}
\begin{split}
\partial^2_{\sigma}\phi(\sigma,\tau)\simeq&\int_0^{\infty}\frac{dq}{\pi}\Biggl\{\left(1-e^{-2q^4\tau}\right)\left(\frac{q^2}{q^4+x^2}-\frac{1}{q^2}\right)\\
&+2\phi(\sigma,\tau)\left[1+e^{-2q^4\tau}\left(\frac{q^4}{q^4+x^2}-1\right)\right]\\
&-4q^4\int_0^{\tau}d\tau'\,\phi(\sigma,\tau')e^{-2q^4(\tau-\tau')}\Biggr\}.
\end{split}
\end{equation}
The first factor in line one of this equation is of order $\mathcal{O}(q^4\tau)\ll1$ for $q\lesssim\tau^{-1/4}$ and bounded for $q\rightarrow\infty$, whence the entire first line may be neglected. Moreover, the exponential occurring in line two may be set to unity, since the term multiplying this exponential vanishes for wave numbers $q\gtrsim\tau^{-1/4}\gg1$, where this constitutes a bad approximation. We are thus left with
\begin{equation}
\label{eq:pullshort2}
\begin{split}
\partial^2_{\sigma}\phi(\sigma,\tau)&\simeq\int_0^{\infty}\frac{dq}{\pi}\Biggl\{2\phi(\sigma,\tau)\,\frac{q^4}{q^4+x^2}\\
&-4q^4\int_0^{\tau}d\tau'\,\phi(\sigma,\tau')e^{-2q^4(\tau-\tau')}\Biggr\}.
\end{split}
\end{equation}
Finally, mapping Eq. \eqref{eq:pullshort2} to Laplace space and using $x\ll1$, we eventually arrive at the following linearized asymptotic differential equation
\begin{equation}
\label{eq:pulllaplace}
\partial^2_{\sigma}\bar{\phi}(\sigma,z)=2^{-3/4}z^{1/4}\bar{\phi}(\sigma,z),
\end{equation}
which is to be solved subject to the boundary conditions (cf. Eqs. \eqref{eq:BC})
\begin{subequations}
\label{eq:BCpulllaplace}
\begin{eqnarray}
\bar{\phi}(0,z) &=& z^{-2}\\
\lim_{\sigma\rightarrow\infty}\bar{\phi}(\sigma,z) &=& 0.
\end{eqnarray}
\end{subequations}
In Eq. \eqref{eq:pulllaplace} $\bar{\phi}(\sigma,z)\equiv\int_0^{\infty}d\tau\,e^{-z\tau}\phi(\sigma,\tau)$ denotes the Laplace transform of $\phi(\sigma,\tau)$. One easily verifies, that
\begin{equation}
\label{eq:pulllaplacesolutionshort}
\bar{\phi}(\sigma,z)=z^{-2}e^{-2^{-3/8}z^{1/8}\sigma}
\end{equation}
is the solution to the boundary value problem stated in Eqs. \eqref{eq:pulllaplace} and \eqref{eq:BCpulllaplace}. The real space solution, following from Eq. \eqref{eq:pulllaplacesolutionshort} by means of a Laplace back-transform
\begin{equation}
\label{eq:pullsolutionshort}
\begin{split}
\phi(\sigma,\tau)&=\frac{1}{2\pi i}\int_{\epsilon-i\infty}^{\epsilon+i\infty}\frac{dz}{z^2}e^{-2^{-3/8}z^{1/8}\sigma+z\tau}\\
&=\tau\cdot\frac{1}{2\pi i}\int_{\epsilon'-i\infty}^{\epsilon'+i\infty}\frac{dz}{z^2}e^{-2^{-3/8}z^{1/8}\frac{\sigma}{\tau^{1/8}}+z}\\
&\equiv\tau\cdot\hat{\phi}\left(\xi=\frac{\sigma}{\tau^{1/8}}\right),
\end{split}
\end{equation}
confirms our scaling assumption Eq. \eqref{eq:scalingformphi} with $\alpha=0$ and $\eta=1/8$. Here we substituted $z\tau\rightarrow z$ in the second line. Identifying 
\begin{equation}
\label{eq:xilpa}
\xi=\frac{\sigma}{\tau^{\eta}}\equiv\frac{s}{\lpa(t)},
\end{equation}
we immediately read off 
\begin{equation}
\label{eq:lpapullshort}
\lpa(t)\simeq\sqrt{\frac{\ell_p}{\zet}}\,t^{1/8},
\end{equation}
corroborating our previous scaling analysis. As already discussed on a scaling level in section \ref{sec:s}, the short time asymptote of Eq. \eqref{eq:PIDErescaled} (with $a=0$ and $b=x^2$) is identical to the scenario of unconfined pulling. We thus skip the explicit calculation of $\phi(\sigma,\tau)$ in Eq. \eqref{eq:pullsolutionshort}, referring the interested reader to Ref. \cite{Hallatschek:2007p229}, where a more detailed discussion of this function, including a summation formula for $\phi$, is given.

\subsubsection{\label{sec:pulllong}Long times}
In order to discuss the situation for long times $\mathfrak{f}^2t\gg\tau\gg1$, we split the right hand side of Eq. \eqref{eq:PIDErescaled} into a deterministic ($D$) and a stochastic ($S$) contribution. Adjusting the parameters $a$ and $b$ to the case of pulling, we write
\begin{subequations}
\label{eq:DSpull}
\begin{eqnarray}
D &\equiv& \int_0^{\infty}\frac{dq}{\pi}\frac{q^2\left(1-e^{-2q^2\left[q^2\tau+\phi(\sigma,\tau)\right]-2x^2\tau}\right)}{q^4+x^2}\\
S &\equiv& -\int_0^{\infty}\frac{dq}{\pi}\,2q^2\int_0^{\tau}d\tilde{\tau}\times\\
  &      & e^{-2q^2\left[q^2(\tau-\tilde{\tau})+\phi(\sigma,\tau)-\phi(\sigma,\tilde{\tau})\right]-2x^2(\tau-\tilde{\tau})}\nonumber.
\end{eqnarray}
\end{subequations}
Starting with the stochastic contribution $S$, we first apply a quasi-static approximation, assuming the chain to be equilibrated under the current local backbone tension \cite{BrochardWyart:1999p798}. This technically amounts to an expansion of the integrated tension $\phi(\sigma,\tilde{\tau})$ in the exponential about $\tilde{\tau}=\tau$, \textit{i.e.} approximating $\phi(\tau)-\phi(\tilde{\tau})\approx\partial_{\tau}\phi(\tau)(\tau-\tilde{\tau})$. Invoking our general scaling assumptions \eqref{eq:scalingformphi} and \eqref{eq:alpha} this strategy is rigorously justifiable \cite{Hallatschek:2007p229}. As a result, we are left with
\begin{equation}
\label{eq:Sapproxpullint}
S\simeq-\int_0^{\infty}\frac{dq}{\pi}\frac{1-e^{-2\left[q^2\partial_{\tau}\phi(\sigma,\tau)+x^2\right]\tau}}{q^2+\partial_{\tau}\phi(\sigma,\tau)}
\end{equation}
Here we omitted the term $x^2\ll1$ in the denominator and invoked dominance of tension modes to neglect the bending modes in the exponential. Eq. \eqref{eq:Sapproxpullint} may even be simplified further. To this end we note that the function $(q^2+\partial_{\tau}\phi)^{-1}$ decays over a distance $\Delta q=\mathcal{O}(\tau^{\alpha/2})$, whereas the function $(1-e^{-2[\dots]\tau})$ approaches unity within a much smaller interval $\delta q=\mathcal{O}(\tau^{-(\alpha+1)/2})\ll\Delta q$. We may therefore set the latter function to unity and find
\begin{equation}
\label{eq:Sapproxpull}
S\simeq-\int_0^{\infty}\frac{dq}{\pi}\frac{1}{q^2+\partial_{\tau}\phi(\sigma,\tau)}=\frac{-1}{2\sqrt{\partial_{\tau}\phi(\sigma,\tau)}}.
\end{equation}
In contrast to the stochastic contribution $S$, whose asymptotic form \eqref{eq:Sapproxpull} follows for all times $\tau\gg1$ from a quasi-static approximation, the deterministic contribution $D$ possesses two distinct asymptotes in the long-time regime. To see this, we make use of dominance of tension modes for $\tau\gg1$ to write
\begin{equation}
\label{eq:Dapproxpullint}
D\simeq\int_0^{\infty}\frac{dq}{\pi}\frac{q^2\left(1-e^{-2q^2\phi(\sigma,\tau)-2x^2\tau}\right)}{q^4+x^2}.
\end{equation}
Now the factor $(1-e^{-2q^2\phi-2x^2\tau})$ may be set to unity for all wave numbers $q\gtrsim q_*=\tau^{-(\alpha+1)/2}$, whereas the rational function it multiplies reaches its maximum at $q_m=x^{1/2}$. We thus have to keep the factor $(1-e^{-2q^2\phi-2x^2\tau})$ if $q_m\ll q_*$,\textit{i.e.} if $t\ll\gamma^{-1/(2\alpha+2)}\mathfrak{f}^{-(2\alpha+1)/(\alpha+1)}$, but may omit it once the opposite is true. Anticipating $\alpha=0$ (see below), we thus arrive at
\begin{equation}
\label{eq:Dapproxpull}
D\simeq\int_0^{\infty}\frac{dq}{\pi}
\begin{cases}
\frac{\left(1-e^{-2q^2\phi(\sigma,\tau)}\right)}{q^2} & \mathfrak{f}^{-2}\ll t\ll\frac{1}{\sqrt{\gamma\mathfrak{f}^2}}\\
\frac{1}{q^2+x^2} & \frac{1}{\sqrt{\gamma\mathfrak{f}^2}}\ll t\ll\tlp,
\end{cases}
\end{equation}
where we have to keep the small term $x^2$ in the second line in order to avoid artificial divergencies for $q\rightarrow0$. Note, moreover, that $x^2\tau=\gamma t\ll1$ for $t\ll(\gamma\mathfrak{f}^2)^{-1/2}$ and $\gamma\ll\mathfrak{f}^2$, enabling us to approximate $e^{-2x^2\tau}\approx1$ in the first line. Performing the integrals in Eq. \eqref{eq:Dapproxpull} we obtain
\begin{equation}
\label{eq:Dapproxpull2}
D\simeq
\begin{cases}
\sqrt{\frac{2\phi(\sigma,\tau)}{\pi}} & \mathfrak{f}^{-2}\ll t\ll\frac{1}{\sqrt{\gamma\mathfrak{f}^2}}\\
\frac{1}{2\sqrt{2x}} & \frac{1}{\sqrt{\gamma\mathfrak{f}^2}}\ll t\ll\tlp.
\end{cases}
\end{equation}
As is obvious from physical considerations, the actual tension along the contour is comparable to the externally applied force, whence $\partial_{\tau}\phi=\mathcal{O}(1)$. This implies $S=\mathcal{O}(1)$ and the deterministic term ($D=\mathcal{O}(\tau^{(\alpha+1)/2})\gg1$ and $D=\mathcal{O}(x^{-1/2})\gg1$ respectively) clearly dominates the dynamics in both intermediate asymptotic regimes. Thus, neglecting stochastic contributions and setting 
\begin{equation}
\label{eq:intEMpulllong}
\partial^2_{\sigma}\phi(\sigma,\tau)\simeq D,
\end{equation}
we eventually arrive at the sought-after asymptotic equations of motion.

Here the intermediate asymptote $D\propto\phi^{1/2}$ reproduces the ``taut string approximation'', first used in Ref. \cite{Seifert:1996p779} on a heuristic level in order to discuss tension propagation for large longitudinal pulling forces in free space. In the second long time asymptote, the deterministic contribution $D\propto x^{-1/2}$ is nothing but the initial stored length density (in rescaled units) for polymers equilibrated under zero tension in cylindrical confinements of constant strength $\gamma$. Hence, the corresponding equation of motion might have been readily inferred from our scaling picture of section \ref{sec:s} by assuming the initially stored length to be pulled out completely by virtue of a strong pulling force $\mathfrak{f}$.

As anticipated in section \ref{sec:s}, the equation of motion governing the tension dynamics in the regime $\mathfrak{f}^{-2}\ll t\ll(\gamma\mathfrak{f}^2)^{-1/2}$, where the chain yet has locally no perception of the confining channel walls, is just the same as the one found in Ref. \cite{Hallatschek:2007p229} for unconfined pulling and asymptotically long times. There the corresponding differential equation is discussed in detail. We simply state the solution:
\begin{equation}
\label{eq:pullsoltgg11}
\phi(\sigma,\tau)=\tau\left[\left(\frac{1}{72\pi}\right)^{1/4}\xi-1\right]^4_{\xi=\sigma/\tau^{1/4}},
\end{equation}
again corroborating our scaling assumption with $\alpha=0$ and $\eta=1/4$ and hence
\begin{equation}
\label{eq:lpapulllong1}
\lpa(t)\simeq\sqrt{\frac{\ell_p}{\zet}}\,(\mathfrak{f}t)^{1/4}.
\end{equation}

The second asymptote in Eqs. \eqref{eq:Dapproxpull2} and \eqref{eq:intEMpulllong} may be written down in a parameter free form by means of a redefinition of the dimensionless arc length variable $\sigma$. Setting
\begin{equation}
\label{eq:redefsigma}
x^{-1/4}\sigma\rightarrow\sigma
\end{equation}
and invoking our scaling assumption \eqref{eq:scalingformphi} with $\eta=1/2$ and $\alpha=0$ we find
\begin{equation}
\label{eq:intEMpulllong2}
\partial^2_{\xi}\hat{\phi}(\xi)\simeq\frac{1}{2\sqrt{2}},
\end{equation}
implying a parabolic $\xi$-dependence for $\hat{\phi}$. Noting that the actual backbone tension is given by $f=\mathfrak{f}\,\hat{\varphi}$, where $\hat{\varphi}=\hat{\phi}-\xi\partial_{\xi}\hat{\phi}/2$, we see that Eq. \eqref{eq:intEMpulllong2} predicts a linearly decreasing tension profile. Obviously, this result is meaningful only within the boundary sections of the chain, since a linearly decreasing tension profile cannot smoothly match the boundary condition in the bulk Eq. \eqref{eq:BC2}. Of course, our above reasoning leading to Eq. \eqref{eq:intEMpulllong} was based on the assumption $\varphi=\partial_{\tau}\phi=\mathcal{O}(1)$, which certainly breaks down far away from the boundaries, where the local backbone tension $\varphi\ll1$ is close to zero and stochastic effects dominate the physics. A more complete description thus has to include the influence of thermal motion in order to account for the smooth transition between a linearly decreasing tension within the boundaries and a constant (zero) tension in the bulk \cite{Obermayer:2007p643}. Since the details of this transition, however, are dispensable in the context of longitudinal dynamics, we will not enter this discussion and confine ourselves to the simple picture suggested by Eq. \eqref{eq:intEMpulllong2}, predicting a piecewise linear tension profile. A comparison of this piecewise linear estimate to a full numerical solution of Eq. \eqref{eq:PIDErescaled}, reveals that this is indeed a feasible approximation.
Using the scaling exponent $\eta=1/2$ and recalling the redefinition of $\sigma$ in Eq. \eqref{eq:redefsigma} we infer
\begin{equation}
\label{eq:lpapulllong2}
\lpa(t)\simeq\sqrt{\frac{\ell_p}{\zet}}\,\gamma^{1/8}(\mathfrak{f}t)^{1/2},
\end{equation}
as expected from our scaling picture.

\subsubsection{OPT regime}
So far we concentrated on short times $t\ll\tlp$, where the bulk of the polymer is not influenced by the externally applied tension $\mathfrak{f}$, and where we were able to restrict our discussions to formally semi-infinite chains. In contrast, for times $t\gg\tlp$ tension propagated through the whole contour and we have to account for the finite length of the chain in general. We estimate the crossover time $\tlp$ by setting $\lpa(\tlp)\equiv L$. Using Eq. \eqref{eq:lpapulllong2} we find
\begin{equation}
\label{eq:tlppull}
\tlp=\zet\,\frac{L^2}{\ell_p}\,\frac{1}{\gamma^{1/4}\mathfrak{f}}.
\end{equation}
We shall now show, that this time scale is indeed identical to the time $t_*$, marking the crossover to applicability of OPT. To estimate $t_*$ note that OPT predicts a flat tension profile $f_{\text{OPT}}(s)=\mathfrak{f}$, which in our rescaled version of the time integrated tension amounts to $\phi_{\text{OPT}}(\sigma,\tau)=\tau$. In reality the tension spatially varies along the contour and we may write $\phi(\sigma,\tau)=\tau+\delta\phi(\sigma,\tau)$ for times $t=\mathcal{O}(t_*)$ \cite{Obermayer:2007p643}. Redefining $\sigma\equiv s/L$ and using Eq. \eqref{eq:intEMpulllong2}, which is still valid for times $t\gtrsim\tlp$, we find the following boundary value problem governing $\delta\phi$:
\begin{subequations}
\label{eq:deltaphi}
\begin{eqnarray}
\partial^2_{\sigma}\delta\phi(\sigma,\tau) &\simeq& 2^{-3/2}\zet\,\frac{L^2}{\ell_p}\,\frac{\mathfrak{f}}{\gamma^{1/4}}\\
\delta\phi(\sigma,\tau)\Bigr|_{\sigma=0,1} &=& 0.
\end{eqnarray}
\end{subequations}
The solution to this problem may readily be inferred by straightforward integration. We immediately read off $\delta\phi=\mathcal{O}\left(\zet\,\frac{L^2}{\ell_p}\,\frac{\mathfrak{f}}{\gamma^{1/4}}\right)$. In order for OPT to be applicable we have to require $\delta\phi\ll\phi_{\text{OPT}}=\tau$. From this we deduce the time scale signaling the crossover to OPT
\begin{equation}
\label{eq:tstarpull}
t_*=\zet\,\frac{L^2}{\ell_p}\,\frac{1}{\gamma^{1/4}\mathfrak{f}},
\end{equation}
which turns out to be identical to $\tlp$ stated in Eq. \eqref{eq:tlppull}. Hence we are in a position to invoke OPT in order to determine the longitudinal dynamics in the time regime succeeding tension propagation.

\subsection{Release}
It turns out that confinement effects remain invisible to the relaxing contour during the entire tension propagation regime, due to a large bulk tension $\sim\mathfrak{f}$, which suppresses thermal undulations to amplitudes far below the channel diameter. Within this tension propagation regime, we will therefore confine ourselves to a brief discussion of the approximations necessary to map the equation of motion \eqref{eq:PIDErescaled} (with $a=1$ and $b=x^2$) to the respective asymptotic differential equations, which indeed turn out to be identical to ones discussed in Ref. \cite{Hallatschek:2007p229} in the context of unconfined release. The actual presence of confining channel walls becomes visible not before the tension has essentially relaxed, resulting in a corresponding shift in the  time scale $t_*$, which marks the crossover to OPT (cf. section \ref{sec:releaseOPT}).

\subsubsection{Tension propagation}
The universal short time regimes $\tau\ll1$ may be treated in an analogous fashion as above (cf. section \ref{sec:pullshort}). Invoking dominance of bending modes and linearizing Eq. \eqref{eq:PIDErescaled} in $\phi$ we find
\begin{equation}
\label{eq:releaseshort}
\begin{split}
\partial^2_{\sigma}\phi(\sigma,\tau)\simeq&\int_0^{\infty}\frac{dq}{\pi}\Biggl\{\left(e^{-2q^4\tau}-1\right)\,\frac{1}{q^2\left(q^2+1\right)}\\
&+2\phi(\sigma,\tau)\left[1+e^{-2q^4\tau}\left(\frac{q^2}{q^2+1}-1\right)\right]\\
&-4q^4\int_0^{\tau}d\tau'\,\phi(\sigma,\tau')e^{-2q^4(\tau-\tau')}\Biggr\}.
\end{split}
\end{equation}
Noting that only wave numbers $q\sim\tau^{-1/4}\gg1$ contribute in the first line, we may set $q^2(q^2+1)\approx q^4$. Moreover, the exponential occurring in line two may be set to unity again. Mapping the resulting expression to Laplace space and performing the remaining $q$-integrals we are left with
\begin{equation}
\label{eq:releaseshortEM}
\partial^2_{\sigma}\bar{\phi}(\sigma,z)\simeq\left(2^{-3/4}z^{1/4}-1\right)\bar{\phi}(\sigma,z)-2^{-3/4}z^{-7/4},
\end{equation}
subject to the boundary conditions (cf Eq. \eqref{eq:BC})
\begin{subequations}
\label{eq:BCreleaselaplace}
\begin{eqnarray}
\bar{\phi}(0,z) &=& 0 \\
\lim_{\sigma\rightarrow\infty}\bar{\phi}(\sigma,z) &=& z^{-2}.
\end{eqnarray}
\end{subequations}
For an explicit treatment of this particular boundary value problem in the short time limit $z\gg1$ we refer the interested reader to Ref. \cite{Hallatschek:2007p229}. There, again, our previously proposed scaling form Eq. \eqref{eq:scalingformphi} is confirmed with $\alpha=0$ and $\eta=1/8$ giving rise to the universal short time scaling law
\begin{equation}
\label{eq:lpareleaseshort}
\lpa(t)\simeq\sqrt{\frac{\ell_p}{\zet}}\,t^{1/8},
\end{equation}
in accordance with our scaling picture.

The asymptote for long times $\tau\gg1$ may be inferred along exactly the same line of reasoning as outlined for the pulling scenario, giving ($x\ll1$)
\begin{equation}
\label{eq:releaselongint}
\partial^2_{\sigma}\phi\simeq\int_0^{\infty}\frac{dq}{\pi}\left\{\frac{1-e^{-2q^2\phi}}{q^2+1}-\frac{1-e^{-2q^2\partial_{\tau}\phi\cdot\tau}}{q^2+\partial_{\tau}\phi}\right\}.
\end{equation}
Moreover, using $\partial_{\tau}\phi=\mathcal{O}(1)$, the rational functions $(q^2+1)^{-1}$ and $(q^2+\partial_{\tau}\phi)^{-1}$ decay over distances $\Delta q=\mathcal{O}(1)$. The exponentials, in contrast, decay over much smaller distances $\delta q=\mathcal{O}(\tau^{-(\alpha+1)/2})\ll1$. It is thus legitimate to neglect the exponentials altogether, to arrive at the following asymptotic differential equation
\begin{equation}
\label{eq:asympdgereleaselong}
\partial^2_{\sigma}\phi(\sigma,\tau)=\frac{1}{2}-\frac{1}{2\sqrt{\partial_{\tau}\phi(\sigma,\tau)}}
\end{equation}
governing the system's tension dynamics in the long time regime. Once again, choosing $\alpha=0$ and $\eta=1/2$, this equations may be solved by means of a one parameter scaling function of type Eq. \eqref{eq:scalingformphi}, giving in particular \cite{Hallatschek:2007p229}
\begin{equation}
\label{eq:lpareleaselong}
\lpa(t)\simeq\sqrt{\frac{\ell_p}{\zet}}\,\mathfrak{f}^{3/4}t^{1/2}.
\end{equation}

\subsubsection{\label{sec:releaseOPT}Homogeneous tension relaxation and crossover to OPT}
Unlike in the case of pulling, the time scales $\tlp$ and $t_*$ are separated by a regime of \textit{homogeneous tension relaxation} \cite{Hallatschek:2007p229} in the release scenario. In particular, setting $\lpa(\tlp)=L$ we find
\begin{equation}
\label{eq:tlprelease}
\tlp=\zet\,\frac{L^2}{\ell_p}\,\mathfrak{f}^{-3/2},
\end{equation}
which is identical to the respective crossover scale observed in unconfined release \cite{Hallatschek:2007p229}.
In contrast to its unconfined counterpart, however, the crossover to applicability of OPT, where the backbone tension has essentially relaxed to its new equilibrium value $f(s,t)=0$, occurs significantly earlier since confinement supersedes relaxation of modes $q\lesssim\gamma^{1/4}$. To estimate the time scale $t_*$ marking the transition to OPT we calculate the change in stored length predicted by OPT and compare the corresponding friction force $\zet L^2\langle\Delta\rho\rangle_{\text{OPT}}/t$ to the characteristic force $\lp^{-2}$ of the currently relaxing mode. Noting that the right hand side of Eq. \eqref{eq:PIDErescaled} is proportional to the change in stored length (cf. Eq. \eqref{eq:tensionstoredlength}), we obtain (setting $\phi=0$)
\begin{equation}
\label{eq:drhoOPTrelease}
\begin{split}
\langle\Delta\rho\rangle_{\text{OPT}} &\simeq \frac{-1}{\ell_p\mathfrak{f}^{1/2}}\int_0^{\infty}\frac{dq}{\pi}\left\{\frac{1}{q^2+1}-\frac{q^2}{q^4+x^2}\right\}\\
                                      &=\mathcal{O}\left(\frac{\gamma^{-1/4}}{\ell_p}\right),
\end{split}
\end{equation}
which could have been anticipated by means of equilibrium theory. In deriving this result we used $x\ll1$ and omitted the factors $\left(1-e^{-2(q^4+x^2)\tau}\right)$, which approach unity within a fairly small interval $\delta q=\mathcal{O}(\tau^{-1/4})\ll1$ and thus do not affect the integrals. Comparison of friction and characteristic force $\lp^{-2}\sim t^{-1/2}$ (chain relaxes freely, \textit{i.e.} under vanishing tension) thus provides us with the crossover time scale
\begin{equation}
\label{eq:tstarrelease}
t_*=\zet^2\frac{L^4}{\ell_p^2}\,\gamma^{-1/2},
\end{equation}
which---as anticipated---is much smaller than the respective time scale for unconfined chains, provided the confinement is sufficiently strong (cf. section \ref{sec:l}). Given that $x\ll1$ and $\mathfrak{f}>f_c$ the time $t_*$ is indeed large compared to the crossover scale $\tlp$
\begin{equation}
\frac{t_*}{\tlp}\sim\left(\frac{\mathfrak{f}}{f_c}\right)^{1/2}x^{-1}\gg1.
\end{equation}
Here
\begin{equation}
\label{eq:fc}
f_c\equiv\frac{\ell_p^2}{\zet^2L^4}
\end{equation}
denotes a critical scale for the externally applied force, below which the chain reacts linearly to the external perturbation, never entering any non-linear long-time regimes. Focusing on the more interesting cases of non-linear response we tacitly assume the inequality $\mathfrak{f}>f_c$ to be fulfilled. Under these conditions a regime of homogeneous tension relaxation emerges for times $\tlp\ll t\ll t_*$, whose dynamics, governed by the same equation of motion \eqref{eq:asympdgereleaselong}, is identical to the one within the respective unconfined scenario, where the tension is shown to decay $\sim t^{-2/3}$ \cite{Hallatschek:2007p229}.

\subsection{Free relaxation from confinement (FRC)}
As a third scenario we shall investigate the intermediate asymptotics in the case of an initially confined contour relaxing in free space. While strongly increasing the confinement strength eventually leads to creation of hairpin structures, analogous to the case of compressing a polymer \cite{Ranjith:2002p1174}, and thus forbids the use of the presently applied theory for late times, a sudden drop in confinement strength is devoid of such peculiarities and may be described successfully by means of Eq. \eqref{eq:PIDErescaled}. Here we restrict ourselves to the discussion of a sudden change in confinement strength from some finite value $\gamma$ to zero. This scenario may be regarded as another typical example of initially straight contours relaxing in free space \cite{Obermayer:2009p740}. We will show that this problem is indeed very similar to the unconfined release scenario \cite{Hallatschek:2007p229,Obermayer:2009p740}, safe for a transient short time regime, where tension builds up along the contour and the longitudinal dynamics $\dRpa$ will be shown to follow a new superlinear scaling law. In addition, the prefactor for $\dRpa$ in the long time tension propagation regime turns out to be different, due to a somewhat different condition in the bulk.

As a key result of the present subsection we will determine the scaling exponent quantifying the growth of the backbone tension at short times. A quantitative discussion of the longitudinal response $\dRpa$ will be given in section \ref{sec:l}.

\subsubsection{Short times}
For short times ($\tau\ll1$) we again linearize Eq. \eqref{eq:PIDErescaled} (with $a=0$ and $b=1$):
\begin{equation}
\label{eq:confoffshort}
\begin{split}
\partial^2_{\sigma}\phi(\sigma,\tau)\simeq&\int_0^{\infty}\frac{dq}{\pi}\Biggl\{\frac{e^{-2q^4\tau}-1}{q^2(q^4+1)}\\
&+2\phi(\sigma,\tau)\left[1+e^{-2q^4\tau}\left(\frac{q^4}{q^4+1}-1\right)\right]\\
&-4q^4\int_0^{\tau}d\tau'\,\phi(\sigma,\tau')e^{-2q^4(\tau-\tau')}\Biggr\}.
\end{split}
\end{equation}
Applying the same reasoning as before, we set the exponential in line two equal to one and subsequently Laplace transform the resulting expression. Performing the remaining $q$-integrals and focusing on short times, \textit{i.e.} $z\gg1$, we arrive at
\begin{equation}
\label{eq:confoffEMshort}
\partial^2_{\sigma}\bar{\phi}(\sigma,z)\simeq2^{-3/4}z^{1/4}\bar{\phi}(\sigma,z)+2^{-1/2}z^{-2},
\end{equation}
which is subject to the boundary conditions (cf. Eq. \eqref{eq:BC})
\begin{subequations}
\label{eq:BCconfofflaplace}
\begin{eqnarray}
\bar{\phi}(0,z) &=& 0\\
\lim_{\sigma\rightarrow\infty}\partial_{\sigma}\bar{\phi}(\sigma,z) &=& 0.
\end{eqnarray}
\end{subequations}
This linear second order boundary value problem is solved straightforwardly by standard methods yielding
\begin{equation}
\label{eq:solutionconfoffshortlaplace}
\bar{\phi}(\sigma,z)=2^{1/4}z^{-9/4}\left(1-e^{-2^{-3/8}z^{1/8}\sigma}\right).
\end{equation}
From this, the sought-after solution $\phi(\sigma,\tau)$ is found by means of an inverse Laplace transform. Using the linearity of the inverse Laplace transform, as well as 
\begin{equation}
\label{eq:invlaplaceaux}
\mathcal{L}^{-1}\left[z^{-9/4+n/8}\right]=\tau^{5/4}\,\frac{\tau^{-n/8}}{\Gamma\left(\frac{9}{4}-\frac{n}{8}\right)},
\end{equation}
where $n\neq18+8k\;(k=1,2,\dots)$ and $\mathcal{L}^{-1}[.]$ denotes the inverse Laplace transform, we may write down $\phi(\sigma,\tau)$ in the form of a summation formula:
\begin{equation}
\label{eq:solutionconfoffshort}
\begin{split}
\phi(\sigma,\tau)&=2^{1/4}\mathcal{L}^{-1}\left[z^{-9/4}\left(1-e^{-2^{3/8}z^{1/8}\sigma}\right)\right]\\
                 &=2^{1/4}\sum_{n=1}^{\infty}\frac{(-1)^{n+1}2^{-3n/8}\sigma^n}{n!}\,\mathcal{L}^{-1}\left[z^{-9/4+n/8}\right]\\
                 &=2^{1/4}\tau^{5/4}\sum_{n=1}^{\infty}\frac{(-1)^{n+1}2^{-3n/8}}{n!\,\Gamma\left(\frac{9}{4}-\frac{n}{8}\right)}\,\xi^n\Biggr|_{\xi=\sigma/\tau^{1/8}}.
\end{split}
\end{equation}
Here we expanded the exponential in the first line and exclude the values $n=18+8k\;(k=1,2,\dots)$ in the summation range. This result corroborates our general scaling assumption Eq. \eqref{eq:scalingformphi} with $\alpha=1/4$ and $\eta=1/8$. The latter exponent again yields the known universal short time scaling law $\lpa\sim t^{1/8}$. The asymptotic form for the rescaled bulk tension $\varphi_{\infty}\equiv f_{\infty}/\mathfrak{f}$ may readily be inferred from the bulk limit of Eq. \eqref{eq:solutionconfoffshortlaplace}. Noting that $\bar{\varphi}=z\bar{\phi}$ we conclude $\bar{\varphi}_{\infty}=\lim_{\sigma\rightarrow\infty}z\bar{\phi}=2^{1/4}z^{-5/4}$. Thence, using Eq. \eqref{eq:invlaplaceaux} we find
\begin{equation}
\label{eq:bulktensionconfoffshort}
\varphi_{\infty}(\tau)=\frac{2^{1/4}}{\Gamma\left(\frac{5}{4}\right)}\,\tau^{1/4}.
\end{equation}
The backbone tension thus grows $\sim t^{1/4}$ for short times.

\subsubsection{Long times, homogeneous tension relaxation and OPT}
Following our approximation schemes explained for the two force cases Eq. \eqref{eq:PIDErescaled} reduces to
\begin{equation}
\label{eq:confoffasymplong}
\begin{split}
\partial^2_{\sigma}\phi(\sigma,\tau)&\simeq\int_0^{\infty}\frac{dq}{\pi}\left\{\frac{q^2}{q^4+1}-\frac{1}{q^2+\partial_{\tau}\phi(\sigma,\tau)}\right\}\\
                                    &=\frac{1}{2^{3/2}}-\frac{1}{2\sqrt{\partial_{\tau}\phi(\sigma,\tau)}}
\end{split}
\end{equation}
for long times $\tau\gg1$. Taking the bulk limit of Eq. \eqref{eq:confoffasymplong} and using Eq. \eqref{eq:BC2} supplies us with the rescaled bulk tension. We find
\begin{equation}
\label{eq:bulktensionconfofflong}
0=\frac{1}{2^{3/2}}-\frac{1}{2\sqrt{\varphi_{\infty}(\tau)}}\;\Rightarrow\;\varphi_{\infty}=2,
\end{equation}
where we used $\lim_{\sigma\rightarrow\infty}\partial_{\tau}\phi(\sigma,\tau)=\varphi_{\infty}(\tau)$. The bulk tension thus saturates at a constant value
\begin{equation}
f_{\infty}=2\,\gamma^{1/2}
\end{equation}
for asymptotically long times during tension propagation.
Note that the asymptotic equation of motion for the rescaled tension $\varphi=\partial_{\tau}\phi$, entailed by Eq. \eqref{eq:confoffasymplong}, is identical to the respective equation for unconfined release, safe for a shift in boundary conditions: in the present case the rescaled tension $\varphi$ approaches the bulk value $\varphi_{\infty}=2$ rather than $\varphi_{\infty}=1$. This in turn gives rise to a new prefactor in the scaling law for $\dRpa$, which we determined by means of a shooting method (cf. table \ref{tab:dRpa1} and section \ref{sec:l}). Again the scaling for $\lpa(t)$ is obtained by mapping Eq. \eqref{eq:confoffasymplong} to an ordinary differential equation by means of our general scaling assumption \eqref{eq:scalingformphi} and identifying $\xi=s/\lpa(t)$. Choosing $\alpha=0$ and $\eta=1/2$ we find
\begin{equation}
\label{eq:lpaconfofflong}
\lpa(t)\simeq\sqrt{\frac{\ell_p}{\zet}}\,\gamma^{3/8}t^{1/2},
\end{equation}
which is the well known result from unconfined release \cite{Hallatschek:2007p229} (identifying $\gamma^{1/2}\leftrightarrow\mathfrak{f}$).

As soon as the boundary layers span over the whole contour and we have to take into account the finiteness of the chain. The mathematics describing the polymer dynamics now becomes strictly identical to the unconfined release scenario, enabling us to carry over the corresponding results within the regimes of homogeneous tension relaxation and OPT. This in particular gives the crossover time scales:
\begin{subequations}
\begin{eqnarray}
\label{eq:crossovertimesconfoff}
\tlp &=& \zet\,\frac{L^2}{\ell_p}\,\gamma^{-3/4}\\
t_*  &=& \frac{L^8}{\ell_p^4},
\end{eqnarray}
\end{subequations}
which are again separated provided the effective confinement strength is sufficiently strong.

\subsection{Pulling and release for small forces}
For small forces $\mathfrak{f}\ll\gamma^{1/2}$, \textit{i.e.} $x\gg1$, pulling and release become mutually inverse scenarios. Since the respective short time asymptotes remain unchanged we restrict our discussions to asymptotically long times $t\gg\gamma^{-1}$ during tension propagation. The equation of motion \eqref{eq:PIDE} in the case of pulling reads
\begin{equation}
\label{eq:EMxgg1}
\begin{split}
&\partial^2_sF(s,t)\simeq\\
&\simeq\frac{\zet}{\ell_p}\int_0^{\infty}\frac{dq}{\pi}\left\{\frac{q^2}{q^4+\gamma}-\frac{q^2}{q^4+q^2\partial_tF(s,t)+\gamma}\right\}\\
&=\gamma^{-1/4}\frac{\zet}{\ell_p}\int_0^{\infty}\frac{dq}{\pi}\left\{\frac{q^2}{q^4+1}-\frac{q^2}{q^4+q^2\frac{\partial_tF(s,t)}{\gamma^{1/2}}+1}\right\},
\end{split}
\end{equation}
where we invoked a quasi-static approximation (which is justified at any rate, since $\gamma t\gg1$) and where we rescaled wave numbers $q\rightarrow q\gamma^{1/4}$. Note that, in order to describe the release case, we would simply have to add the term $q^2x^{-1}$ in the denominator of the first term. However, since this term contributes only for wave numbers $q\gtrsim x^{1/2}$ where the complete first term is close to zero, it is justified to neglect this contribution altogether, whence pulling and release are both described by means of the same Eq. \eqref{eq:EMxgg1}. Noting that $\gamma^{-1/2}\partial_tF(s,t)=\mathcal{O}(x^{-1})\ll1$, we may expand the integrand in Eq. \eqref{eq:EMxgg1} and find the following diffusion equation for the time integrated tension
\begin{equation}
\label{eq:EMxgg1approx}
\partial_tF(s,t)\simeq8\sqrt{2}\frac{\gamma^{3/4}\ell_p}{\zet}\,\partial^2_sF(s,t)
\end{equation}
with diffusion constant
\begin{equation}
\label{eq:diffusionconstant}
\Delta_0\propto\frac{\gamma^{3/4}\ell_p}{\zet}\sim D^{-2},
\end{equation}
in our units proportional to the inverse cross-sectional area of the channel ($D$ denoting the channel diameter). Here we used Odijk's scaling law $\gamma^{-3/4}\sim L_d^3\sim D^2\ell_p$ \cite{Odijk:1983p211}. From this we immediately deduce the scaling for the boundary layer size $\lpa$ \cite{Nam:2010p1691}:
\begin{equation}
\label{eq:lpxgg1}
\lpa\simeq\sqrt{\frac{\ell_p}{\zet}}\,\gamma^{3/8}t^{1/2}\propto\sqrt{\Delta_0\,t}.
\end{equation}
Once again, Eq. \eqref{eq:EMxgg1approx} may be solved by means of a one parameter scaling function. Defining
\begin{subequations}
\begin{eqnarray}
\label{eq:sigmadefxgg1}
\sigma &\equiv& \frac{\mathfrak{f}}{\sqrt{8\sqrt{2}\Delta_0}}\,s\\
\tau   &\equiv& \mathfrak{f}^{2}\,t
\end{eqnarray}
\end{subequations}
and setting $F(s,t)\equiv\mathfrak{f}^{-1}\phi(\sigma,\tau)$ we obtain
\begin{equation}
\label{EMxgg1resc}
\partial_{\tau}\phi(\sigma,\tau)=\partial^2_{\sigma}\phi(\sigma,\tau).
\end{equation}
Or, using our general scaling assumption Eq. \eqref{eq:scalingformphi} with $\alpha=0$ and $\eta=1/2$ we get
\begin{equation}
\label{EMxgg1oneparam}
\hat{\phi}''(\xi)+\frac{\xi}{2}\hat{\phi}'(\xi)=0.
\end{equation}
Hence, invoking the respective boundary conditions Eqs. \eqref{eq:BC}, we find the following one parameter solution for pulling:
\begin{equation}
\label{eq:solxgg1pull}
\hat{\phi}(\xi)=1-\text{erf}\left(\frac{\xi}{2}\right),
\end{equation}
and for release:
\begin{equation}
\label{eq:solxgg1release}
\hat{\phi}(\xi)=\text{erf}\left(\frac{\xi}{2}\right).
\end{equation}
Here $\text{erf}(\xi)$ denotes the error function. 

For sufficiently weak forces, the relaxation dynamics directly crosses over to the OPT regime in both cases. To estimate the corresponding crossover time $t_*$ it is thus sufficient to set $\lpa(t_*)\equiv L$ giving
\begin{equation}
\label{eq:tstarxgg1}
t_*=\frac{L^2}{\ell_p}\gamma^{-3/4}.
\end{equation}

\section{\label{sec:l}Longitudinal response}

\begin{table*}
	\begin{ruledtabular}
		\begin{tabular}{c|c c c}
Time Regime				&		Pulling		
									&		Release		
									&		Free Relaxation From Confinement		\\
\hline
short times				&		$ [2^{5/8}/\Gamma(15/8)](\mathfrak{f}/\sqrt{\zet\ell_p})\,t^{7/8}$
									&		$-[2^{5/8}/\Gamma(15/8)](\mathfrak{f}/\sqrt{\zet\ell_p})\,t^{7/8}$
									&		$-[2^{7/8}/\Gamma(17/8)](\gamma^{3/4}/\sqrt{\zet\ell_p})\,t^{9/8}$		\\
intermediate
times							&		$[8/(72\pi)^{1/4}](\mathfrak{f}^{3/4}/\sqrt{\zet\ell_p})\,t^{3/4}$
									&		$-$
									&		$-$		\\
long times				&		$ 2^{3/4}(\mathfrak{f}^{1/2}\gamma^{-1/8}/\sqrt{\zet\ell_p})\,t^{1/2}$
									&		$-2.477(\mathfrak{f}^{1/4}/\sqrt{\zet\ell_p})\,t^{1/2}$
		 							&		$-2.946(\gamma^{1/8}/\sqrt{\zet\ell_p})\,t^{1/2}$		\\
$\tlp\ll t\ll t_*$&		$-$
									&		$-18^{1/3}[L/(\zet\ell_p^2)]^{1/3}t^{1/3}$
									&		$-18^{1/3}[L/(\zet\ell_p^2)]^{1/3}t^{1/3}$		\\
$t\gg t_*$				&		$ 2^{-3/2}(\gamma^{-1/4}L/\ell_p)$
									&		$-2^{-3/2}(\gamma^{-1/4}L/\ell_p)$
									&		$-[2^{3/4}/\Gamma(1/4)](L/\ell_p)\,t^{1/4}$
		\end{tabular}
	\end{ruledtabular}
	\caption{Response of the projected end-to-end distance $\dRpa(t)$ for large forces ($x\ll1$). The numerical prefactor for the scenario of free relaxation from confinement (long times) was newly determined by means of a shooting method. The prefactor for pulling (long times) was determined analytically on the basis of Eq. \eqref{eq:intEMpulllong2} by requiring the time integrated tension to be continuously differentiable at the transition to the bulk.}
	\label{tab:dRpa1}
\end{table*}

\begin{table}[b]
	\begin{ruledtabular}
		\begin{tabular}{c| c}
Time Regime				&		Pulling / Release ($x\gg1$)\\
\hline
$t\ll\gamma^{-1}$	&		$\pm[2^{5/8}/\Gamma(15/8)](\mathfrak{f}/\sqrt{\zet\ell_p})\,t^{7/8}$\\
$\gamma^{-1}\ll
t\ll\tlp$					&		$\pm[2^{-3/4}/\sqrt{\pi}](\mathfrak{f}\gamma^{-3/8}/\sqrt{\zet\ell_p})\,t^{1/2}$		\\
$\tlp=t_*\ll t$		&		$\pm(2^{-7/2}L/\ell_p)\mathfrak{f}/\gamma^{3/4}$		\\
		\end{tabular}
	\end{ruledtabular}
	\caption{Response of the projected end-to-end distance $\dRpa(t)$ for the pulling and release scenario in strong confinement for weak forces ($x\gg1$). The corresponding scaling laws only differ by a sign and are linear in the externally applied force $\mathfrak{f}$.}
	\label{tab:dRpa2}
\end{table}

\subsection{Tension propagation and homogeneous tension relaxation}
Up to now our quantitative discussions have focused on the analysis of Eq. \eqref{eq:PIDErescaled} and the derivation of respective scaling solutions for the tension profiles. These results set the stage for an analytical discussion of the longitudinal response of the chain, represented by the change in the projected end-to-end distance $\dRpa(t)$. This quantity is formally expressed in terms of the stored length density:
\begin{equation}
\label{eq:dRpastoredlength}
\begin{split}
\dRpa(t)&=\int_0^Lds\left\langle r_{\parallel}'(s,0)-r_{\parallel}'(s,t)\right\rangle\\
        &=-\int_0^Lds\drho(s,t)+o(\epsilon).
\end{split}
\end{equation}
Omitting terms $o(\epsilon)$ and using Eq. \eqref{eq:tensionstoredlength} we may thus relate longitudinal response and (time integrated) tension profile
\begin{equation}
\label{eq:dRpatension}
\begin{split}
\dRpa(t) &= \frac{1}{\zet}\int_0^Lds\,\partial^2_sF(s,t)=\frac{-2}{\zet}\partial_sF(0,t)\\
				 &=-2\partial_{\xi}\hat{\phi}(\xi=0)\,\frac{\mathfrak{f}^{5/4+2(\alpha-\eta)}}{\sqrt{\zet\ell_p}}\,t^{\alpha+1-\eta},
\end{split}
\end{equation}
where we used the symmetry of the boundary conditions and explicitly inserted our scaling assumption Eq. \eqref{eq:scalingformphi}. Recall that we have to replace $\mathfrak{f}\rightarrow\gamma^{1/2}$ in the FRC scenario. Thus knowledge of the previously determined dynamical exponents $\alpha$ and $\eta$ immediately supplies us with the scaling laws governing the longitudinal response $\dRpa$ within the respective asymptotic time regimes. The only unknown quantity $\partial_{\xi}\hat{\phi}(\xi=0)$ can be determined from the corresponding equations of motion. 
We determined this numerical constant explicitly for the various scenarios and time regimes discussed in this work, either analytically on the basis of the respective asymptotic equation of motion or numerically by means of a shooting method.

\subsection{Ordinary perturbation theory (OPT)}
Within the OPT regime, where the tension profiles are flat, the longitudinal response is calculated directly from change in stored length density.
In the following we give a brief derivation of the respective scaling laws. To this end we calculate the change in stored length density $\langle\Delta\rho\rangle(t)$, which is proportional the right hand side of Eq. \eqref{eq:PIDE}. Within OPT this expression considerably simplifies due to the spatially constant tension profiles. Restoring original units and starting with the pulling scenario we find
\begin{equation}
\label{eq:pullOPTdrho}
\begin{split}
&\langle\Delta\rho\rangle_{\text{OPT}}(t)\simeq\\
&\simeq-\int_0^{\infty}\frac{dq}{\pi\ell_p}\left(1-e^{-2q^2\mathfrak{f}\,t}\right)\left\{\frac{q^2}{q^4+\gamma}-\frac{q^2}{q^4+q^2\mathfrak{f}+\gamma}\right\}\\
&\approx-\int_0^{\infty}\frac{dq}{\pi\ell_p}\left\{\frac{q^2}{q^4+\gamma}-\frac{q^2}{q^4+q^2\mathfrak{f}+\gamma}\right\}\\
&\approx-\frac{1}{2^{3/2}}\,\frac{\gamma^{-1/4}}{\ell_p}.
\end{split}
\end{equation}
Here we invoked dominance of tension modes and eventually neglected the exponential in the second line, which decays on a scale of the order $\mathcal{O}((\mathfrak{f}\,t)^{-1/2})$ which is much smaller than the scale characteristic of the rational functions ($\mathcal{O}(\gamma^{1/4})$) for times $t\gg\gamma^{-1/2}\mathfrak{f}^{-1}$. Thence, noting that $x\ll1$, we find the change in stored length density to be identical to the (negative) stored length density of a polymer equilibrated in a constant confinement of effective strength $\gamma$, corroborating our scaling picture drawn in section \ref{sec:s}. Finally, using $\dRpa=-\langle\Delta\rho\rangle\,L$ we arrive at the scaling law given in table \ref{tab:dRpa1}. 

For the FRC scenario we have to keep the exponential in the stochastic term in order to avoid artificial divergencies for $q\rightarrow0$. We find
\begin{equation}
\begin{split}
\langle\Delta\rho\rangle_{\text{OPT}}(t)&\simeq-\int_0^{\infty}\frac{dq}{\pi\ell_p}\left\{\frac{q^2}{q^4+\gamma}-\frac{1-e^{-2q^4t}}{q^2}\right\}\\
                           &\approx\frac{2^{3/4}}{\ell_p\,\Gamma\left(\frac{1}{4}\right)}\,t^{1/4},
\end{split}
\end{equation}
where we used $t^{1/4}\gg\gamma^{-1/4}$.

Finally, the change in stored length density for the release scenario was already calculated in Eq. \eqref{eq:drhoOPTrelease}.

We conclude this section by briefly discussing the case of large force scale separation parameter $x\gg1$ in the context of pulling and release. Note that within OPT, pulling and release are mutually ``reverse'' scenarios (cf. table \ref{tab:dRpa1}). Using equation \eqref{eq:pullOPTdrho} and rescaling wave numbers ($q\rightarrow q\,\gamma^{1/4}$) we find
\begin{equation}
\label{eq:pullreleaseOPTdrhoxgg1}
\begin{split}
\langle\Delta\rho\rangle(t)&\simeq\pm\gamma^{-1/4}\int_0^{\infty}\frac{dq}{\pi\ell_p}\left\{\frac{q^2}{q^4+x^{-1}q^2+1}-\frac{q^2}{q^4+1}\right\}\\
                           &=\mp\gamma^{-1/4}x^{-1}\int_0^{\infty}\frac{dq}{\pi\ell_p}\frac{q^4}{\left(q^4+1\right)^2}+\mathcal{O}\left(x^{-2}\right)\\
                           &\approx\mp\frac{2^{-7/2}}{\ell_p}\,\frac{\mathfrak{f}}{\gamma^{3/4}},
\end{split}
\end{equation}
where we expanded the first line in $x^{-1}\ll1$. In Eq. \eqref{eq:pullreleaseOPTdrhoxgg1} the upper sign refers to pulling, the lower to release.

Our results are summarized in tables \ref{tab:dRpa1} and \ref{tab:dRpa2}. Here, as already discussed in the previous section, the force scenarios are identical to their unconfined counterparts for sufficiently short times \cite{Hallatschek:2007p229}, whereas the dynamical scaling relations are analogous to the cases of prestretched chains for late times \cite{Obermayer:2007p643}, consistent with our earlier conclusion to interpret confinement as effective prestretching force. Unlike the respective free space scenarios, however, the force scenarios in confinement saturate at a constant value for $\dRpa$ within the OPT regime $t\gg t_*$, which is due to the suppression of long wavelength modes by confinement.
The FRC scenario bears close resemblance to the scenario of unconfined release for late times $t\gg\gamma^{-1}$, safe for a shift in the prefactor of $\dRpa$ which is due to a somewhat different boundary condition in the bulk (cf. previous section). In particular the universal $\dRpa\sim t^{1/3}$ scaling, characteristic of freely relaxing ``initially straight'' contours (cf. Ref. \cite{Obermayer:2009p740}), is recovered. Most remarkably, the FRC scenario exhibits a superlinear $\dRpa\sim t^{9/8}$ scaling for short times, which may be traced back to a steadily growing tension in the bulk.

\subsection{\label{sec:sublim}Weak confinement}
Our hitherto discussions for large external forces ($x\ll1$) revealed that the two force scenarios are governed by the dynamical laws of the respective unconfined problems for sufficiently short times. In the limit $\gamma\rightarrow0$, of course, all the confinement induces differences vanish and we be able recover the results of Ref. \cite{Hallatschek:2007p229}. We will now show that there exists a finite critical value of the confinement strength $\gamma$, below which confinement effects are irrelevant for the longitudinal dynamics.

\subsubsection{Pulling}
The pulling scenario starts to deviate from its unconfined counterpart as soon as the equilibration length $\lp(t)$ grows beyond Odijk's deflection length $\gamma^{1/4}$ at times $t\gtrsim\gamma^{-1/2}\mathfrak{f}$. Subsequently the chain's dynamics is governed by scaling laws which closely resemble those of prestretched polymers \cite{Obermayer:2007p643} and have no analog for unconfined molecules without prestress. In contrast to unconfined chains the longitudinal dynamics saturates at a constant value of the projected end-to-end distance for times beyond $t_*$, which is thus identical to the longest relaxation time.

In order for these observations to be true, however, we tacitly assumed the following ordering in time to hold: $\mathfrak{f}^{-2}\ll\gamma^{-1/2}\mathfrak{f}^{-1}\ll t_*$, where $t_*$ is given in Eq. \eqref{eq:tstarpull}. While the first inequality is true by virtue of a large force separation $x\ll1$, the second inequality imposes a constraint on the confinement strength $\gamma$. More precisely we have to stipulate
\begin{equation}
\label{eq:critgammapull}
\gamma\gg\gamma_c\equiv f_c^{2}.
\end{equation}
The critical force $f_c$ (Eq. \eqref{eq:fc}) thus translates into a critical confinement strength $\gamma_c$, below which the third asymptotic time regime within tension propagation vanishes. Violation of Eq. \eqref{eq:critgammapull} has even further consequences. First of all, since the scaling law $\lpa\sim (\mathfrak{f}\,t)^{1/4}$ (rather than $\lpa\sim (\mathfrak{f}\,t)^{1/2}$) governs the growth of $\lpa$ at the crossover between tension propagation and OPT regimes, Eq. \eqref{eq:tstarpull} has to be replaced in favor of 
\begin{equation}
\label{eq:tstarpull<}
t_{*,<}=\zet^2\,\frac{L^4}{\ell_p^2}\,\frac{1}{\mathfrak{f}},
\end{equation}
which is identical to the crossover scale in unconfined pulling. The longitudinal dynamics within OPT also changes. Note that neglecting the exponential in Eq. \eqref{eq:pullOPTdrho} is feasible only if the exponential approaches zero sufficiently fast, \textit{i.e.} only if $(\mathfrak{f}\,t)^{1/2}\sim(\mathfrak{f}\,t_{*,<})^{1/2}\ll\gamma^{1/4}$. For $\gamma<\gamma_c$ this condition is violated and we need to reevaluate Eq. \eqref{eq:pullOPTdrho}:
\begin{equation}
\label{eq:pullOPTdrho<}
\begin{split}
&\langle\Delta\rho\rangle_{\text{OPT}}(t)\simeq\\
&\simeq-\int_0^{\infty}\frac{dq}{\pi\ell_p}\left(1-e^{-2q^2\mathfrak{f}\,t}\right)\left\{\frac{q^2}{q^4+\gamma}-\frac{q^2}{q^4+q^2\mathfrak{f}+\gamma}\right\}\\
&\approx-\sqrt{\frac{2}{\pi}}\,\frac{\sqrt{\mathfrak{f}\,t}}{\ell_p},
\end{split}
\end{equation}
where we neglected the second term in the integrand, which is small compared to the first one, given $x\ll1$. This is---again---identical to the unconfined case. In summary confinement effects vanish once $\gamma$ falls below the critical confinement threshold $\gamma_c$.

\subsubsection{Release}
We will now show that the same conclusions carry over to the case of release. Due to large local values of the backbone tension, the presence of confining channel walls merely affects the dynamical characteristics of the release scenario within the OPT regime, where the projected end-to-end distance saturates at a constant value. Moreover, just like in the case of pulling, sufficiently strong confinement shifts the crossover time scale $t_*$ below its unconfined counterpart such that the crossover to OPT (and---simultaneously---even complete relaxation of the chain) occurs at significantly earlier times.

Again, all these conclusions are implicitly based on the assumption $\gamma\gg\gamma_c$. To see how confinement effects fade away for $\gamma\ll\gamma_c$, note that the validity of Eq. \eqref{eq:drhoOPTrelease} is based on the assumption $\gamma t\gg1$, which is reasonable for $t=\mathcal{O}(t_*)$ ($t_*$ given in Eq. \eqref{eq:tstarrelease}) provided $\gamma\gg\gamma_c$. For $\gamma\ll\gamma_c$, in contrast, we have to keep the neglected exponential in the second term \footnote{The rational function in the first term varies on a scale of order $\mathcal{O}(\sqrt{\mathfrak{f}})$, which is large compared to the scale on which the exponential approaches zero ($\mathcal{O}(t^{-1/4})$), provided $t\gg\mathfrak{f}^{-2}$, which we alway assume to hold (the opposite case corresponds to small forces $\mathfrak{f}<f_c$, where the chain reacts linearly to the externally applied force).} of Eq. \eqref{eq:drhoOPTrelease}, which then becomes
\begin{equation}
\label{eq:drhoOPTrelease<}
\begin{split}
&\langle\Delta\rho\rangle_{\text{OPT}}\simeq\\
&\simeq-\int_0^{\infty}\frac{dq}{\pi\ell_p}\left\{\frac{1}{q^2+\mathfrak{f}}-\frac{q^2\left(1-e^{-2q^4t}\right)}{q^4+\gamma}\right\}\\
&\approx\int_0^{\infty}\frac{dq}{\pi\ell_p}\frac{1-e^{-2q^4t}}{q^2}=\frac{2^{3/4}}{\ell_p\Gamma\left(\frac{1}{4}\right)}\,t^{1/4}.
\end{split}
\end{equation}
Here we neglected $\gamma$ in the denominator of the second term, which affects the corresponding rational function only for wave numbers $q\lesssim\gamma^{1/4}$ where the term $\left(1-e^{-2q^4t}\right)$ is essentially zero. Moreover we neglected the first term in line two, which is of order $\mathcal{O}(\mathfrak{f}^{-1/2})$ and therefore negligible compared to the second term, which is of order $\mathcal{O}(t^{1/4})$, for times $t\gg\mathfrak{f}^{-2}$. Eq. \eqref{eq:drhoOPTrelease<} is again identical to the unconfined release scenario, whence also the time scale $t_*$, which was determined in Eq. \eqref{eq:tstarrelease} on the basis of $\langle\Delta\rho\rangle_{\text{OPT}}$, has to be replaced by 
\begin{equation}
\label{eq:tstarrelease<}
t_{*,<}=\frac{\zet^4L^8}{\ell_p^4},
\end{equation}
which is again identical to the one determined in Ref. \cite{Hallatschek:2007p229}. In order to be consistent, this time scale has to be much smaller than $\gamma^{-1}$, which again yields the condition $\gamma\ll\gamma_c$.

\section{\label{sec:experiments}Contact with experiments}
In order to make contact with experiments, this section presents the experimental time, force and length scales implied by our analytical findings. To this end we need to translate between effective confinement strength $\gamma$ and actual channel diameter $D$.
Such a mapping can be performed by comparing the mean square transverse fluctuations of chains confined in a hard-walled channel to those trapped in harmonic confinements. This procedure yields
\begin{equation}
\label{eq:gammaD}
\gamma=\frac{1}{64\alpha_{\circ}^4}\,\ell_p^{-4/3}D^{-8/3},
\end{equation}
where $\alpha_{\circ}\approx0.17$ has been determined by Yang \textit{et al.} \cite{Yang:2007p1148} by means of Monte Carlo simulations. Note that Eq. \eqref{eq:gammaD} is an equilibrium estimate, which we will extrapolate to non-equilibrium situations in what follows.

We used this relation to exemplarily calculate typical time scales for DNA and F-actin molecules in prototypical experimental setups for the three scenarios discussed in this work, cf. table \ref{tab:timescales}. Note, however, that the characteristic time scales strongly depend on the adjustable parameters $L$, $\gamma$ and $\mathfrak{f}$ whence the respective time windows may be varied considerably in actual experiments.

Eq. \eqref{eq:gammaD} moreover allows us to translate the critical confinement strength $\gamma_c$ into a critical channel diameter $D_c$, above which confinement effects become invisible in the context of tension propagation and longitudinal response. Recalling $\gamma_c=f_c^2$ and using Eqs. \eqref{eq:gammaD} and \eqref{eq:fc} we find:
\begin{equation}
\label{eq:Dc}
D_c=2^{-9/4}\left(\frac{\zet}{\alpha_{\circ}}\right)^{3/2}\,\frac{L^3}{\ell_p^2}.
\end{equation}
For DNA molecules in typical experimental setups, $D<D_c$ is automatically fulfilled due to their small persistence lengths. Considering the much stiffer actin filaments (using $L\approx13.5\,\mu\text{m}$ and $\ell_p\approx16\,\mu\text{m}$ \cite{LeGoff:2002p271}) we find $D_c\approx10\,\mu\text{m}$, such that confinement effects are only visible in rather narrow channels in this case.

Finally, Eq. \eqref{eq:gammaD} may be utilized to investigate the accessibility of the limits $x\gg1$ and $x\ll1$. Note that $f_{\gamma}\equiv\kappa\,\gamma^{1/2}=k_BT/2\,\mu^{-2/3}\ell_p^{1/3}D^{-4/3}$, where $f_{\gamma}$ gives the effective confinement force in original units.
The forces applied in single molecule experiments are of order $\mathfrak{f}\sim1\,\text{pN}$ \cite{Bustamante:2003p870}. Regarding F-actin as telling example ($\ell_{p,\text{actin}}\approx16\,\mu\text{m}$ \cite{Ott:1993p1483,LeGoff:2002p271}), channel diameters between $D=2\,\mu\text{m}$ and $D=10\,\mu\text{m}$ have been used in experiments on confined F-actin molecules \cite{Koster:2005p943}, giving rise to small force scale separation parameters of order $x\sim10^{-2}$ to $x\sim10^{-3}$. On the other hand, channel diameters down to $D=30\,\text{nm}$ \cite{Reisner:2005p1004} and forces as small as $\mathfrak{f}\sim10^{-2}\,\text{pN}$ \cite{Smith:1992p871} have been used in experiments on DNA. Using $\ell_{p,\text{DNA}}\approx50\,\text{nm}$ \cite{Bustamante:2003p870,Reisner:2005p1004} we obtain $x\sim10^2$ in this case. Both limits $x\ll1$ and $x\gg1$ are thus accessible in experiments.

\begin{table}[t]
	\begin{ruledtabular}
		\begin{tabular}{c| c c}
Time Scale				&		F-Actin & DNA\\
\hline
Pulling: & & \\
$\mathfrak{f}^{-2}$	            &		$10^{-4}\,\text{s}$ & $10^{-7}\,\text{s}$\\
$1/(\gamma^{1/2}\mathfrak{f})$	&		$0.01\,\text{s}$    & $10^{-6}\,\text{s}$\\
$\tlp$	                        &		$0.03\,\text{s}$    & $0.09\,\text{s}$\\
\hline
Release: & & \\
$\mathfrak{f}^{-2}$	&		$10^{-4}\,\text{s}$    & $10^{-7}\,\text{s}$\\
$\tlp$	            &		$0.003\,\text{s}$    & $0.02\,\text{s}$\\
\hline
FRC: & & \\
$\gamma^{-1}$	&		$1.2\,\text{s}$    & $10^{-5}\,\text{s}$\\
$\tlp$	      &		$3.5\,\text{s}$    & $1.2\,\text{s}$\\
		\end{tabular}
	\end{ruledtabular}
	\caption{Characteristic time scales for the scenarios discussed in this work. For F-actin chains we assumed $\ell_p=16\,\mu$m, $L=13.5\,\mu$m $D=2\,\mu$m and $\mathfrak{f}=2\,$pN; for DNA we used $\ell_p=50\,$nm, $L=18.6\,\mu$m $D=100\,$nm and $\mathfrak{f}=2\,$pN.}
	\label{tab:timescales}
\end{table}

\section{\label{sec:su}Summary}

In summary we have presented a detailed discussion of confinement effects on the relaxation dynamics of weakly bending polymers. Our work was based on the investigation of three paradigmatic scenarios including a sudden application and release of a point force $\mathfrak{f}$ at the polymer's ends (\textit{confined pulling} and \textit{confined release}), as well as \textit{free relaxation from confinement (FRC)}.

In the context of tension propagation and longitudinal response we showed that for both force scenarios dynamical signatures resulting from confinements become visible only above a certain threshold for the effective confinement strength $\gamma$, which is given by $\gamma_c=\ell_p^4/(\zet^4L^8)$. This translates into an upper channel diameter (cf. Eq. \eqref{eq:Dc}) which imposes almost no restrictions for experiments on DNA but requires channel widths below 
$10\,\mu$m for actin in order to observe effects caused by confinement. For weak confinement strengths the chain's dynamics reduces to that of unconfined chains. For sufficiently strong confinement, we introduced the force scale separation parameter $x=\gamma^{1/2}/\mathfrak{f}$ in order to quantify the strength of confinement relative to the externally applied force $\mathfrak{f}$. We showed that both cases $x\gg1$ and $x\ll1$ are accessible in experiments and consequently derived the chain's dynamics for comparably strong ($x\ll1$) and weak ($x\gg1$) forces $\mathfrak{f}$. The scaling laws governing the longitudinal dynamics of confined chains in the two force scenarios turned out to bear close resemblance with those of prestretched polymers \cite{Obermayer:2009p740}, suggesting to view confinement as effective prestretching force of strength $\sim\gamma^{1/2}$ in this context. In particular, the polymer's longitudinal response $\dRpa(t)$ saturates at its new equilibrium value already for times $t>t_*$ (cf. Eqs. \eqref{eq:tstarpull} and \eqref{eq:tstarrelease}), even though $t_*$ is smaller than the usually obtained value for the longest relaxation time. This behavior is explained by a suppression of long wave lengths in the mode spectrum of the stored length density due to confinement and has already been observed for prestretched chains where it was termed ``premature saturation'' \cite{Obermayer:2007p643}.

The interpretation of confinement as effective prestretching force carries over to the scenario of free relaxation from confinement (FRC), which is for late times identical to the release scenario discussed in Ref. \cite{Hallatschek:2007p229}, safe for a different prefactor of $\dRpa$ which arises as a consequence of somewhat different bulk conditions. For short times, however, the initially tension free contour builds up backbone tension in the bulk, which grows like $\sim t^{1/4}$ before it saturates at the constant value $2\gamma^{1/2}$. This tension build-up in turn affects the longitudinal dynamics of the chain, which responds superlinearly ($\dRpa\sim t^{9/8}$) in this regime. The FRC scenario thus constitutes an intriguing example of initially straight contours, which were discussed quite generally in Ref. \cite{Obermayer:2009p740}.

\begin{acknowledgments}
We gratefully acknowledge financial support by the Deutsche Forschungsgemeinschaft Contract No. FR~850/8--1 and by the German Excellence Initiative via the program ``Nanosystems Initiative Munich (NIM)''.
\end{acknowledgments}

\end{document}